\documentclass[12pt]{iopart}
\usepackage{iopams}
\usepackage{cite}
\usepackage{soul}
\usepackage{graphicx}
\usepackage{color}\usepackage[colorlinks,linkcolor=blue,urlcolor=blue,citecolor=blue,anchorcolor=blue]{hyperref}
\usepackage{braket}
\newcommand{\llangle}{\langle\!\langle}
\newcommand{\rrangle}{\rangle\!\rangle}
\newcommand{\com}[1]{\textcolor{black}{#1}}
\newcommand{\changed}[1]{\textcolor{black}{#1}}

\usepackage[margin=2.46cm]{geometry}

\begin{document}

\title{Correlated Dephasing Noise in Single-photon Scattering}

\author{Tom\'as Ramos$^{1,2}$ and Juan Jos\'e Garc\'ia-Ripoll$^1$}
\address{$^1$ Instituto de F{\'\i}sica Fundamental IFF-CSIC, Calle Serrano 113b, Madrid 28006, Spain}
\address{$^2$ Centro de \'Optica e Informaci\'on Cu\'antica, Facultad de Ciencias, Universidad Mayor, Chile}

\ead{t.ramos.delrio@gmail.com}
\begin{indented}
\item[]\today
\end{indented}

\begin{abstract}
We \com{develop a theoretical framework to describe} the scattering of photons against a two-level quantum emitter \com{with} arbitrary correlated dephasing noise. This is particularly relevant to waveguide-QED setups with solid-state emitters, such as superconducting qubits or quantum dots, which couple to complex dephasing environments in addition to the propagating photons along the waveguide. Combining input-output theory and stochastic methods, we predict the effect of correlated dephasing in single-photon transmission experiments with weak coherent inputs. \com{We discuss homodyne detection and photon counting of the scattered photons and show that both measurements give the modulus and phase of the single-photon transmittance despite the presence of noise and dissipation.} In addition, we demonstrate that these spectroscopic measurements contain the same information as standard time-resolved Ramsey interferometry, \com{and thus they can be used} to \com{fully} characterize the noise correlations without direct access to the emitter. The method is exemplified with paradigmatic correlated dephasing models such as colored Gaussian noise, white noise, telegraph noise, and 1/f-noise, \com{as typically encountered in solid-state environments.}
\end{abstract}

\section{Introduction}

\begin{figure}[t!]
\center
\includegraphics[width=0.6\linewidth]{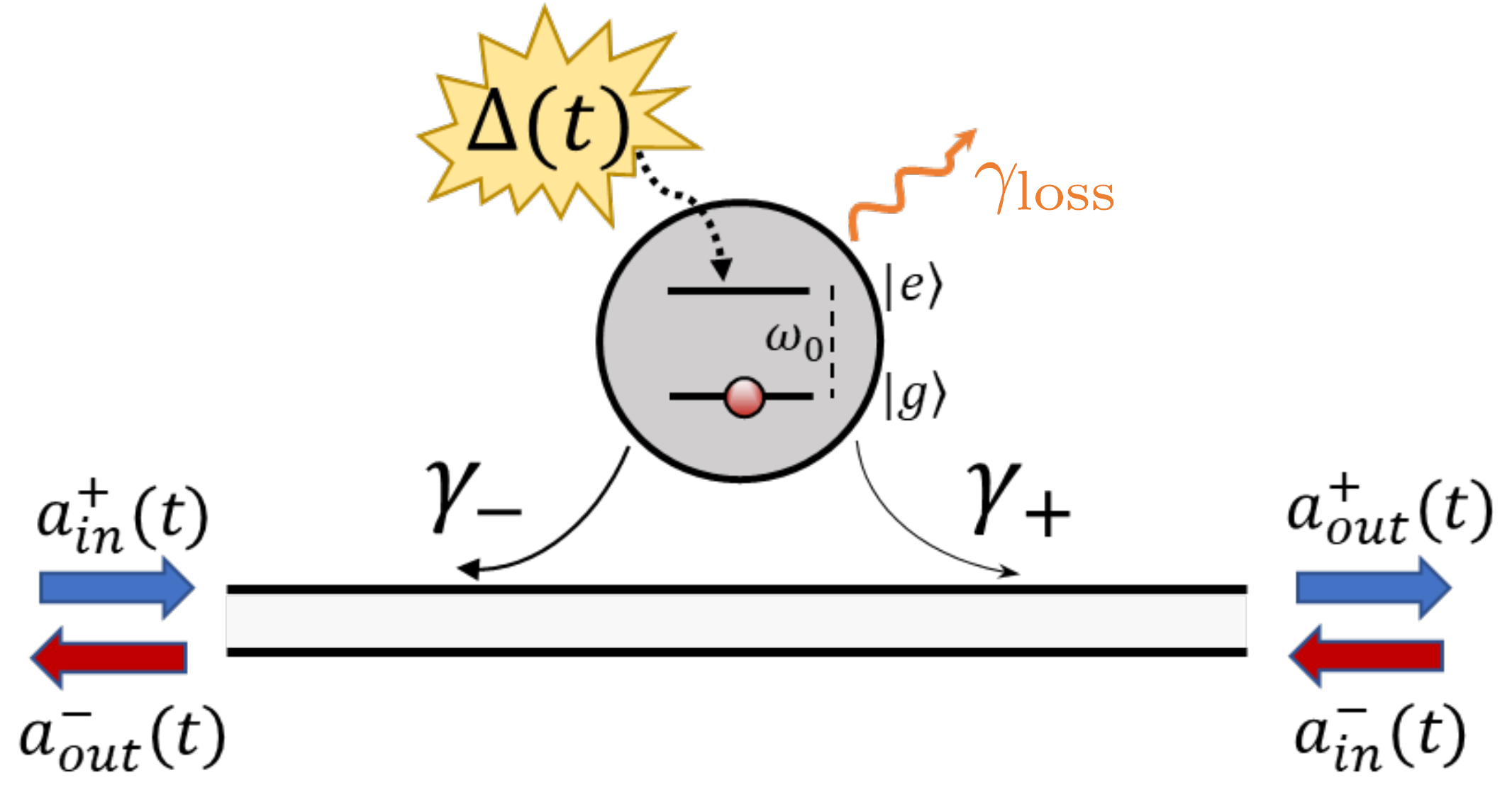}
\caption{A noisy qubit with arbitrary correlated dephasing noise $\Delta(t)$ couples with rates $\gamma_{\pm}$ to right- and left-propagating photons along a 1D waveguide. The qubit also decays with rate $\gamma_{\rm loss}$ into unguided modes. We model dephasing as a stochastic process, and photon scattering via input-output theory with operators $a_{\rm in}^{\pm}(t)$ and $a_{\rm out}^{\pm}(t)$. 
\label{fig:setup}}
\end{figure}

The field of waveguide-QED \cite{roy17,chang18} describes a variety of experimental setups where a quantum emitter interacts preferentially with a family of guided photonic modes, so that the emission rates $\gamma_{\pm}$ into the waveguide approaches or even surpasses decay $\gamma_{\rm loss}$ into unwanted modes (see Figure ~\ref{fig:setup}). This regime \com{has been achieved, for instance,} in experiments with superconducting circuits\ \cite{astafiev10,hoi13,eder18,Gu17}, neutral atoms\ \cite{reitz13,tiecke14,goban15,solano17}\com{, molecules} \cite{wrigge08}, and quantum dots in photonic crystals\ \cite{arcari14,yalla14,coles16}. With a few exceptions, such as \cite{forn-diaz17,magazzu18}, most experiments work in the Rotating-Wave Approximation (RWA) regime, allowing for a adequate description in terms of one- and few-photon wavefunctions\ \cite{shen05,shen07a,zheng10}, input-output theory\ \cite{fan10,caneva15,sanchezburillo16}, \com{diagrammatic methods \cite{roulet16,see17,hurst18},} and path integral formalism\ \cite{shi09,shi15}. Those descriptions usually do not account for other sources of error, such as dephasing, but we know that 1/f-noise severely affects all solid state devices \cite{Revpaladino14}, including quantum dots and superconducting circuits. There have been some experimental attempts at characterizing noise sources \emph{outside} actual circuits, directly exploring the dynamics of the quantum scatterer using time-resolved methods \cite{ithier2005,deppe07,meriles10,Bylander2011,omalley2015,romach15,ethiermajcher17,Norris18,alvarez11,norris16,mavadia18,lisenfeld16} or Fourier transform spectroscopy \cite{kammerer02,berthelot06,coolen08,sallen10,wolters13,thoma16}. Those detailed studies require time-resolved measurements and direct control of the quantum scatterer in many cases, something which may be unfeasible or undesirable in waveguide-QED setups. 

The purpose of this work is to develop a framework of waveguide-QED and scattering theory that accounts for general correlated dephasing, teaching us how to probe a qubit's noise and environment using few-photon scattering experiments. There are earlier works connecting noise with spectroscopy: \changed{Kubo's} fluctuation-dissipation relations links dephasing to lineshapes in nuclear magnetic resonance (NMR)\ \cite{kubo63}, as do later works in the field of quantum chemistry\ \cite{geva97}. Our study complements those works, focusing on the quantum mechanical processes associated to single- and multiphoton scattering in waveguide-QED. We \changed{write down} a stochastic version of the input-output formalism that \com{consistently} includes dephasing noise in the energy levels of the quantum emitter \com{in addition to the dissipative dynamics due to the coupling to photons in the waveguide.} We \changed{then} relate the correlations in that noise to the average scattering matrix of individual photons and coherent wavepackets, and develop strategies to extract those correlations from actual experiments, in conjunction with earlier approaches to scattering tomography\ \cite{ramos17}.

The paper and our main results are organized as follows. In Sec.~\ref{sec:model} we introduce the model for a noisy two-level emitter in a waveguide. We describe dephasing noise as a stationary stochastic process $\Delta(t)$ and derive \changed{the} stochastic input-output equations. In Sec.~\ref{sec:ramsey}, \com{we review the standard procedure of Ramsey interferometry and show how to quantify the noise correlations} via the \textit{Ramsey envelope} $C_\phi(t)$. \com{We also introduce paradigmatic correlated noise models, which will be essential to understand the scattering results in the next sections. In particular,} section~\ref{sec:scattering} shows that the same information provided by Ramsey spectroscopy can be obtained from single-photon scattering experiments, where we only manipulate the qubit through the scattered photons. \changed{We} solve the stochastic input-output equations for a qubit that interacts with a single propagating photon, and show that the \emph{averaged single-photon scattering matrix} can be related one-to-one to the Ramsey envelope. We also discuss analytical predictions for scattering under realistic dephasing models such as \emph{colored Gaussian noise} and \emph{1/f noise}. We show how the noise correlations modify the spectral lineshapes on each case, \changed{recovering} simple limits such as the Lorentzian profiles that are fitted in most waveguide-QED experiments. Section~\ref{sec:coherentextraction} generalizes these ideas, showing how to measure the averaged scattering matrix using weak coherent state inputs together with homodyne or photon \changed{counting} measurements, and how to reconstruct the Ramsey envelope $C_\phi(t)$ from such spectroscopic measurements. This opens the door to the reconstruction of more general correlated noise models that are non-Gaussian but common in many solid-state environments such as telegraph noise and 1/f noise due to ensembles of two-level fluctuators. \changed{We treat this separately} in \ref{sec:NoiseModels} due to the higher complexity of the stochastic methods needed for the analysis. We close this work in Sec.~\ref{sec:summary}, discussing the conclusions and open questions. 

\section{Model for a noisy qubit in a photonic waveguide}\label{sec:model}

Our study considers the setup depicted in Figure \ref{fig:setup}: a two-level quantum emitter or \emph{qubit} \com{is strongly coupled to a 1D photonic waveguide}, emitting photons with rates $\gamma_{\pm}$ along opposite directions, while simultaneously interacting with an \com{unwanted} environment that induces correlated dephasing and dissipation into unguided modes. \com{The Hamiltonian of the total system can be decomposed as}
\com{\begin{eqnarray}
H (t) = H_{\rm qb}(t) + H_{\rm ph} + H_{\rm qb-ph},\label{totalHam}
\end{eqnarray}
and below we describe each term.}

\com{First, the qubit Hamiltonian is given by 
\begin{eqnarray}
H_{\rm qb}(t) = \frac{1}{2}[\omega_0+\Delta(t)]\sigma_z,\label{Hqubit}
\end{eqnarray}
where $\sigma_z=|e\rangle \langle e|-|g\rangle \langle g|$ is the diagonal Pauli operator, with $|e\rangle$ and $|g\rangle$ the excited and ground states of the qubit.} We phenomenologically model environment-induced dephasing as a stochastic fluctuation \com{$\Delta(t)$} of the qubit frequency around a mean value $\omega_0$. We assume the stochastic process $\Delta(t)$ \cite{jacobs2010,vankampen92,gardiner1985} has vanishing mean ---i.e. the \emph{stochastic average} $\llangle\ldots\rrangle$ over noise realizations is zero $\llangle\Delta(t)\rrangle=0$ ---, and  is \emph{stationary} ---i.e. all expectation values and noise correlations $\llangle \Delta(t_1)\dots \Delta(t_n) \rrangle$ are invariant under a global shift in time---. The simplest autocorrelation function $\llangle \Delta(0)\Delta(\tau) \rrangle$ defines a characteristic correlation time $\tau_c$ of the noise as,
\begin{eqnarray}
\tau_c = \int_0^\infty\! d\tau\ \frac{\llangle \Delta(0)\Delta(\tau) \rrangle}{\llangle \Delta^2(0) \rrangle}.
\end{eqnarray}
Those conditions and the machinery of stochastic methods \cite{jacobs2010,vankampen92,gardiner1985} account for any realistic source of qubit dephasing, including arbitrary correlated Markovian and non-Markovian noise, or 1/f noise, among the examples considered below.

\com{The second term in Eq.~(\ref{totalHam}) corresponds to the Hamiltonian of free photons propagating in the waveguide and in unguided modes},
\com{\begin{eqnarray}
H_{\rm ph} = \sum_{\mu=\pm} \int d\omega\ \omega a_{\omega}^{\mu}{}^\dag a_{\omega}^{\mu} + \int d\omega\ \omega b_{\omega}^\dag b_{\omega}.\label{qubitphotonInt}
\end{eqnarray}
Here, the annihilation operator $a_{\omega}^{\mu}$ destroys a photon of frequency $\omega$ propagating to the right $(\mu=+)$ and left $(\mu=-)$ of the waveguide, whereas $b_{\omega}$ destroys an unguided photon of frequency $\omega$. They satisfy standard commutation relations $[a_{\omega}^{\mu},a_{\omega'}^{\mu'}{}^\dag]=\delta_{\mu\mu'}\delta(\omega-\omega')$ and $[b_{\omega},b_{\omega'}^\dag]=\delta(\omega-\omega')$. We consider the RWA throughout this work, so that photons are only populated in a narrow bandwidth around the mean frequency of the qubit $\omega_0$, and the integration limits of $\omega$ in Eq.~(\ref{qubitphotonInt}) can be safely extended to $\pm \infty$ \cite{QuantumNoise,fan10}.} 

\com{The last term in the Hamiltonian (\ref{totalHam}) describes the qubit-photon interaction, which in the RWA reads}
\com{\begin{eqnarray}
H_{\rm qb-ph} = \sum_{\mu=\pm} \sqrt{\frac{\gamma_\mu}{2\pi}}\int d\omega \left(\sigma^+ a^\mu_\omega + {\rm h.c.} \right) + \sqrt{\frac{\gamma_{\rm loss}}{2\pi}} \int d\omega \left(\sigma^+ b_\omega + {\rm h.c.} \right),\label{qbphInt}
\end{eqnarray}
with $\sigma^+=|e\rangle \langle g|$ and $\sigma^-=|g\rangle \langle e|$ the raising and lowering qubit operators. The qubit absorbs and emits photons at a rate $\gamma_\mu$ for waveguide photons in direction $\mu=\pm$, and at a rate $\gamma_{\rm loss}$ for unguided photons. \changed{Assuming a Markov approximation in the qubit-photon coupling ($\gamma_\mu, \gamma_{\rm loss}, |\Delta(t) |\ll \omega_0$)}, the dynamics of the photons can be integrated out, and the noisy qubit is effectively governed by quantum Langevin equations \cite{fan10,gardinercollett85,QuantumNoise}, given in the Heisenberg picture as}
\begin{eqnarray}
\fl\qquad\frac{d \sigma^{-}}{dt}&=&-\left(\frac{\Gamma}{2}+i[\omega_0+\Delta(t)]\right)\sigma^{-}+i\sigma_z\sum_{\mu=\pm}\sqrt{\gamma_\mu}a^{\mu}_{\rm in}(t)+i\sigma_z\sqrt{\gamma_{\rm loss}}b_{\rm in}(t),\label{EqMotSm}\\
\fl\qquad\frac{d\sigma_z}{dt}&=&-\Gamma(\sigma_z+1)-2i\sum_{\mu=\pm}\sqrt{\gamma_\mu}(\sigma^{+}a_{\rm in}^{\mu}(t)-{\rm h.c.})-2i\sqrt{\gamma_{\rm loss}}(\sigma^{+}b_{\rm in}(t)-{\rm h.c.}).\label{EqMotSz}
\end{eqnarray}
Here, the total decay of the qubit $\Gamma=\gamma + \gamma_{\rm loss}$, combines the emission of the qubit into guided $\gamma=\gamma_++\gamma_-$ and unguided modes $\gamma_{\rm loss}$. While typical qubit-waveguide couplings are symmetric $\gamma_{\pm}=\gamma/2$, our formalism with independent channels ($\mu=\pm$) naturally admits the possibility of a \emph{`chiral'} waveguide with different couplings to left- and right-moving photons $\gamma_-\neq \gamma_+$ \cite{roy10,ramos2014,LodahlReview}. \com{The initial condition of the photons is determined via the Heisenberg operators $a_{\rm in}^{\mu}(t)$ and $b_{\rm in}(t)$, which describe the input field photons in the  waveguide and unguided modes, respectively, and read 
\begin{eqnarray}
\fl \hspace{1cm} a_{\rm in}^\mu (t) = \frac{1}{\sqrt{2\pi}} \int d\omega e^{-i\omega (t-t_0)} a_\omega^\mu(t_0),\qquad {\rm and}\qquad b_{\rm in}(t) = \frac{1}{\sqrt{2\pi}} \int d\omega e^{-i\omega (t-t_0)} b_\omega(t_0),
\end{eqnarray}
with $a_\omega^\mu(t_0)$ and $b_\omega(t_0)$ the Heisenberg operators at the initial time.} 

\com{After interacting with the qubit, the photons leave the waveguide through the right $(\mu=+)$ and left $(\mu=-)$ output ports, where they can be measured. The output \changed{fields} of the \changed{waveguide} photons \changed{are} described by the output \changed{operators} $a_{\rm out}^{\mu}(t)$, which \changed{are} given by input-output relations as \cite{fan10,gardinercollett85}}
\begin{eqnarray}
a_{\rm out}^{\mu}(t)&=a_{\rm in}^{\mu}(t)-i\sqrt{\gamma_{\mu}}\sigma^{-}(t),\qquad {\rm with}\ \mu=\pm.\label{InputOutput}
\end{eqnarray}
\com{These equations allow us to access the information of the qubit's dynamics via the \changed{waveguide} photons and will be essential for the optical characterization of the correlated dephasing noise}.

Due to the random classical field $\Delta(t)$, the equations of motion (\ref{EqMotSm}) and (\ref{EqMotSz}) are \emph{stochastic} differential equations. In such equations, each particular realization of the noise provides different quantum expectation values $\braket{\sigma^-}$ or $\braket{\sigma_z}$, and we need to average over all possible noise realizations to obtain more meaningful and measurable values ---i.e. $\llangle \braket{\sigma^-}\rrangle$ or $\llangle \braket{\sigma_z}\rrangle$, as well as higher order multi-time correlations if needed---. In the remainder of the paper we calculate this kind of stochastic averages to characterize the effect of correlated dephasing noise on both the time-resolved dynamics (Sec.~\ref{sec:ramsey}) and the single-photon spectroscopy (Secs.~\ref{sec:scattering}-\ref{sec:coherentextraction}) of the qubit.

\section{Time-resolved characterization of correlated dephasing noise}\label{sec:ramsey}

\com{In this section, we first review the concepts of Ramsey interferometry (see Sec.~\ref{ramseyIntro}) and then introduce the paradigmatic model of colored Gaussian noise (see Sec.~\ref{IntrocoloredGaussian}), which can be analytically solved for arbitrary noise correlation times $\tau_c$. Reviewing these concepts will be essential to understand the effect of correlated \changed{dephasing} in the photon scattering of the next sections.} 

\subsection{Ramsey interferometry}\label{ramseyIntro}

\begin{figure}[t!]
\center
\includegraphics[width=\linewidth]{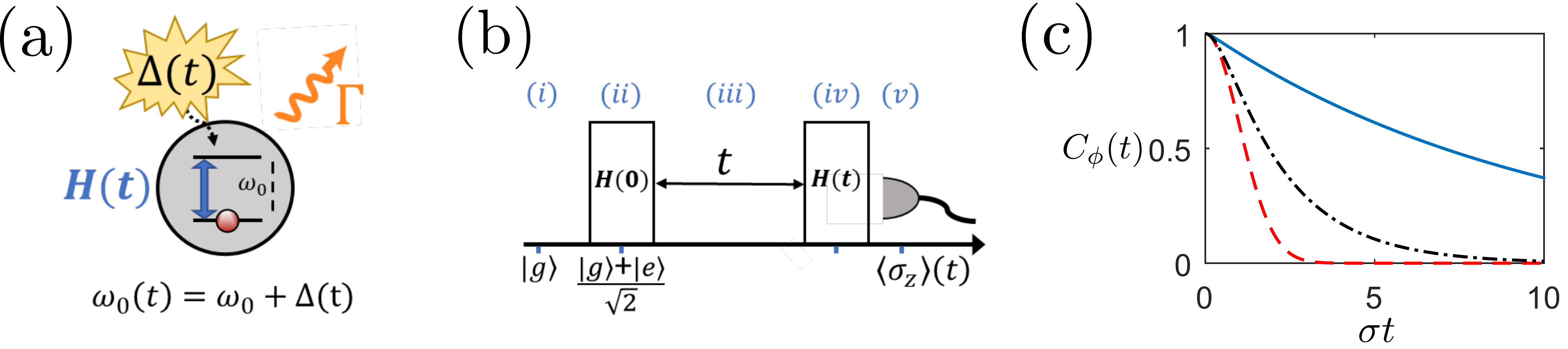}
\caption{Time-resolved characterization of correlated dephasing noise. (a) A qubit coupled to a generic noisy environment causing pure dephasing $\Delta(t)$, and radiative relaxation with rate $\Gamma$. (b) Basic Ramsey sequence consisting of (i) ground state preparation, (ii) Hadamard gate $H(0)$, (iii) free evolution, (iv) second Hadamard gate $H(t)$, and (v) measurement of the qubit population difference $\braket{\sigma_z(t)}$. (c) Ramsey envelopes $C_\phi(t)$ for a noisy qubit with colored Gaussian dephasing, characterized by the noise strength $\sigma$ and the correlation time $\tau_c=1/\kappa$ (see Eq.~(\ref{OUexact})). For $\kappa=10\sigma$ the noise is in the white limit and $C_\phi(t)$ is exponential (blue/solid line). For $\kappa=0$ the noise is quasi-static and $C_\phi(t)$ is Gaussian (red/dashed line). Finally, for $\kappa=2\sigma$ the decay interpolates between \changed{the} two previous behaviors.
\label{fig:ramsey}}
\end{figure}

Ramsey interferometry \cite{ithier2005,deppe07,ethiermajcher17} is the most common way to characterize qubit decoherence. This and other time-resolved methods require full control and read-out of the qubit, while it is in contact with its environment [see Figure \ref{fig:ramsey}(a)]. These methods are \changed{experimentally demanding}, but give detailed information about noise correlations, specially when combined with dynamical decoupling \cite{Bylander2011,alvarez11} and other control techniques \cite{omalley2015,Norris18,norris16,mavadia18}.

A standard Ramsey sequence consists of the five steps from figure \ref{fig:ramsey}(b): (i) Preparation of the qubit in its ground state $\ket{g}$, (ii) application of a Hadamard gate $H(0)$ with a very fast $\pi/2$ pulse, (iii) evolution of the qubit for a time $t$, (iv) application of a second Hadamard gate $H(t)$, and (v) measurement of the qubit population difference $\braket{\sigma_z}$. The purpose of steps (i)-(ii) is to produce the initial superposition state,
\begin{eqnarray}
\ket{\Psi(0)}=H(0)\ket{g}\ket{0}=\frac{1}{\sqrt{2}}(\ket{e}+\ket{g})\ket{0},\label{InitialRamsey}
\end{eqnarray}
for which the qubit coherence is maximal, namely $\braket{\sigma^-(0)}=1/2$. Steps (iii)-(v) monitor the destruction of the qubit coherence $\braket{\sigma^-(t)}$, under the influence the noisy environment. Repeating this procedure for various waiting times $t$ and averaging over many realizations, one obtains the average coherence $\llangle\braket{\sigma^-}\rrangle(t)$.

The dynamics of the qubit coherence under the influence of pure dephasing and radiative decay is obtained by taking expectation values on the quantum Langevin equation (\ref{EqMotSm}). For the initial condition (\ref{InitialRamsey}), it reads
\begin{eqnarray}
\frac{d}{dt}\braket{\sigma^-}&=&-\left(\frac{\Gamma}{2}+i[\omega_0+\Delta(t)]\right)\braket{\sigma^-},\label{EqCoherence}
\end{eqnarray}
which is a multiplicative stochastic differential equation with a random variable $\Delta(t)$\ \cite{jacobs2010,vankampen92,gardiner1985}. To solve for the average $\llangle \braket{\sigma^-} \rrangle (t)$, we integrate equation\ (\ref{EqCoherence}) formally and average the result over all stochastic realizations of the random trajectory $\Delta(t)$, obtaining
\begin{eqnarray}
\llangle \braket{\sigma^-} \rrangle ={}& \frac{1}{2}e^{-(\Gamma/2+i\omega_0)t} C_\phi(t).\label{decayPhi}
\end{eqnarray}
In addition to the exponential decay with rate $\Gamma$ due to the coupling to photons \footnote{\com{Notice that the relaxation decay $\Gamma$ can be obtained independently of the dephashing noise by measuring $\llangle \braket{\sigma_z}\rrangle$ without the second Hadamard gate in Figure \ref{fig:ramsey}, resulting in the pure exponential decay,     
\begin{eqnarray}
\llangle \braket{\sigma_z} \rrangle=e^{-\Gamma t}-1. 
\end{eqnarray}
}}, pure dephasing originates an extra decay factor $C_\phi(t)$ known as Kubo's relaxation function \cite{kubo63} or ``Ramsey envelope'' \cite{omalley2015}. For stationary noise, $C_\phi(t)$ is the average of the random phase accummulated by the qubit after a time $t$, \changed{namely}
\begin{eqnarray}
C_\phi(t)=\llangle e^{-i\int_{0}^t dt' \Delta(t')}\rrangle.\label{DephasingRelaxation}
\end{eqnarray}In general, this function depends on noise correlations of arbitrary order $\llangle \Delta(t_1)\ldots\Delta(t_n) \rrangle$ \changed{whose characterization requires} sophisticated noise spectroscopy methods~\cite{norris16,kotler13}, but for Gaussian noise we will find that only first and second moments are required, as shown below.

\subsection{Colored Gaussian noise, white noise, and quasi-static noise}\label{IntrocoloredGaussian}

For stationary Gaussian noise with vanishing mean, all cummulants and correlations can be expressed in terms of the autocorrelation $\llangle \Delta(0)\Delta(\tau) \rrangle$ \cite{kubo63,vankampen92}, and thus $C_{\phi}(t)$ in Eq.~(\ref{DephasingRelaxation}) is reduced to
\begin{eqnarray}
C_\phi(t)={\rm exp}\left(-\int_0^t d\tau (t-\tau)\llangle \Delta(0)\Delta(\tau) \rrangle\right).\label{gaussianCphi}
\end{eqnarray}
If the noise is also Markovian, Doob's theorem \cite{vankampen92} implies that the noise can be described as an Ornstein-Uhlenbeck process \cite{jacobs2010,vankampen92,gardiner1985} with autocorrelation given explicitly by
\begin{eqnarray}
\llangle \Delta(0)\Delta(\tau)\rrangle=\sigma^2e^{-\kappa|\tau|}.\label{OUcorrelation}
\end{eqnarray}
This rich model describes ``colored Gaussian noise'' with strength $\sigma=\llangle \Delta^2(0)\rrangle^{1/2}$ and a correlation time $\tau_c=1/\kappa$, that covers both fast and slow noise limits. This includes \emph{white noise} with autocorrelation $\llangle \Delta(0)\Delta(\tau) \rrangle=2\gamma_\phi \delta(\tau)$, when taking the \com{limits $\kappa\rightarrow\infty$ and $\sigma\rightarrow\infty$, while keeping a constant pure dephasing rate $\gamma_\phi=\sigma^2/\kappa$}. It also includes \emph{quasi-static noise} in the opposite limit of $\kappa \rightarrow 0$, in which  the autocorrelation becomes constant $\llangle \Delta(0)\Delta(\tau) \rrangle=\sigma^2$.

Another advantage of the colored Gaussian noise (\ref{OUcorrelation}) is that the Ramsey envelope (\ref{gaussianCphi}) can be derived analytically
\begin{eqnarray}
C_{\phi}(t)={\rm exp}\left(-(\sigma/\kappa)^2(e^{-\kappa t}+\kappa t -1)\right).\label{OUexact}
\end{eqnarray}
This super-exponential envelope has been fitted to experimental data to quantify the strength and correlation of realistic environments\ \cite{ithier2005,omalley2015}. Figure \ref{fig:ramsey}(c) shows the typical shape of this envelope in the the limits of fast and slow noise. In the white noise limit (blue/solid line), the decay is exponential $C_\phi(t)={\rm exp}(-\gamma_\phi t)$ with $\gamma_\phi=\sigma^2/\kappa$; in the quasi-static limit limit (red/dashed), the decay is Gaussian $C_\phi(t)={\rm exp}(-\sigma^2t^2/2)$; and for intermediate values such as $\kappa=2\sigma$, the curve clearly interpolates between both shapes (black/dotted).

\section{Single-photon scattering from a qubit with correlated dephasing noise}\label{sec:scattering}

In the following we use the \com{stochastic input-output formalism of Sec.~\ref{sec:model}} to compute the average single-photon scattering matrix for a qubit with stationary dephasing noise $\Delta(t)$. \com{Most importantly, in Sec.~\ref{averageSMatrix} we derive the stochastic differential equation to solve for the average transmittance $\llangle t_\omega^\mu \rrangle$ and reflectance $\llangle r_\omega^\mu \rrangle$ in a single-photon scattering experiment.} \com{In the spirit of Kubo \cite{kubo63}, we also show that these scattering coefficients contain the same noise correlations as the Ramsey envelope $C_\phi(t)$, which can be determined by an independent time-resolved experiment as shown in the previous section.} Finally, we evaluate $\llangle t_\omega^\mu \rrangle$ for a qubit with colored Gaussian noise \com{(see Sec.~\ref{LineShapeColoredGaussian}) and 1/f noise (see Sec.~\ref{LineShape1f}), discussing on each case the broadening of the spectral lineshape and the connections to well known results in the literature.}

\subsection{Average single-photon scattering matrix}\label{averageSMatrix}

The \emph{single-photon scattering matrix} $S^{\lambda\mu}_{\nu\omega}$ describes the interaction between an isolated photon and a quantum emitter. It is defined as the probability amplitude for the emitter to transform an incoming photon with frequency $\omega$ in channel $\mu=\pm$ into an outgoing photon with possibly different frequency $\nu$ and direction $\lambda=\pm$:
\begin{eqnarray}
S^{\lambda\mu}_{\nu\omega}&=\langle g|\langle 0| a_{\rm out}^{\lambda}(\nu)a_{\rm in}^{\mu}{}^\dag(\omega)|g\rangle |0\rangle.\label{singlePhotonSDef}
\end{eqnarray}
The monochromatic input-output photonic operators $a_{\rm in}^{\mu}(\omega)$ and $a_{\rm out}^{\mu}(\omega)$ are given by the Fourier transform ${\cal F}$ of the Heisenberg input-output field amplitudes defined above \cite{fan10}:
\begin{eqnarray}
a_{\rm in}^{\mu}(\omega)&={\cal F}\left[a_{\rm in}^{\mu}(t)\right](\omega),\qquad a_{\rm out}^{\mu}(\omega)&={\cal F}\left[a_{\rm out}^{\mu}(t)\right](\omega),\label{monochromaticFromField}
\end{eqnarray}
with ${\cal F}[f(t)](\omega)=(2\pi)^{-1/2}\int_{-\infty}^{\infty} dt e^{i\omega t} f(t)$ for a test function $f(t)$. Notice that these monochromatic operators (\ref{monochromaticFromField}) satisfy canonical bosonic commutation relations as well as their time-domain counterparts, namely $[a_{\rm in}^{\lambda}(\nu), a_{\rm in}^{\mu}{}^\dag(\omega)]=[a_{\rm out}^{\lambda}(\nu), a_{\rm out}^{\mu}{}^\dag(\omega)]=\delta_{\lambda\mu}\delta(\nu-\omega)$.

The scattering matrix of the noisy qubit is derived by combining equations (\ref{InputOutput}), (\ref{singlePhotonSDef}), and (\ref{monochromaticFromField}) to obtain
\begin{eqnarray}
S^{\lambda\mu}_{\nu\omega}=\delta_{\lambda\mu}\delta(\nu-\omega)-\sqrt{\frac{\gamma_{\lambda}\gamma_{\mu}}{2\pi}}{\cal F}[G_\omega(t)](\nu-\omega).\label{SgeneralTime}
\end{eqnarray}
Here, the \emph{scattering overlap}, $G_\omega(t)=ie^{i\omega t}(2\pi/\gamma_\mu)^{1/2}\langle 0| \sigma^{-}(t)a^{\mu}_{\rm in}{}^\dag(\omega)|0\rangle$ satisfies an inhomogeneous \changed{stochastic} differential equation derived from Eqs.~(\ref{EqMotSm})-(\ref{EqMotSz}),
\begin{eqnarray}
\frac{dG_\omega}{dt}=&-\left(\frac{\Gamma}{2}-i[\omega-\omega_0]+i\Delta(t)\right)G_{\omega}(t)+1,\label{ScatteringEq}
\end{eqnarray}
\changed{and} with initial condition \com{$G_\omega(t_0)=0$ for $t_0\rightarrow -\infty$}. This is similar to the equation for the qubit's coherence\ (\ref{EqCoherence}), but now including a constant source term. 

As explained in Sec.~\ref{sec:coherentextraction}, spectroscopic measurements are not related to $S$ but to the average scattering matrix $\llangle S^{\lambda\mu}_{\nu\omega}\rrangle$. Computing this quantity is a two-step process. First, we formally integrate equation (\ref{ScatteringEq}) for a stationary noise $\Delta(t)$ \com{and solve for the average $\llangle G_\omega (t)\rrangle$}. Using the stationary noise property $\llangle e^{-i\int_{t-\tau}^t dt' \Delta(t')} \rrangle = \llangle e^{-i\int_{0}^\tau dt' \Delta(t')} \rrangle = C_\phi(\tau)$ \com{and taking the limit $t_0\rightarrow -\infty$, we find that the solution is independent of time, namely}
\com{
\begin{eqnarray}
\llangle G_\omega (t)\rrangle = \llangle G_\omega \rrangle ={\cal L}[C_\phi(\tau)](\Gamma/2-i[\omega-\omega_0]).\label{GLaplace}
\end{eqnarray}}
Note how the noise correlations enter via the Kubo relaxation function $C_\phi(t)$ in Eq.~(\ref{DephasingRelaxation}) after a Laplace transform ${\cal L}[f(t)](s)=\int_{0}^{\infty} dt e^{-st} f(t)$. The second step is to take the \com{stochastic average in Eq.~(\ref{SgeneralTime}) and insert the Fourier transform of Eq.~(\ref{GLaplace}) which} is trivially given by ${\cal F}[\llangle G_{\omega}(t)\rrangle](\nu-\omega)=\sqrt{2\pi}\llangle G_{\omega}\rrangle\delta(\nu-\omega)$. The total averaged scattering matrix then reads
\begin{eqnarray}
\llangle S^{\lambda\mu}_{\nu\omega}\rrangle &=&\left\lbrace\delta_{\lambda\mu}-\sqrt{\gamma_\lambda\gamma_\mu}\llangle G_{\omega}\rrangle\right\rbrace\delta(\nu-\omega),\label{AverageSfinal}
\end{eqnarray} 
\com{where the delta function $\delta(\nu-\omega)$ indicates that the scattering conserves the energy of the photons \emph{on average}. On each realization, we can imagine the emitter absorbing a photon when its transition frequency is $\omega_0+\Delta(t)$, and then relaxing by spontaneous emission when it has a different frequency $\omega_0+\Delta(t')$. During this process, the dephasing environment exerts work on the qubit, adding and subtracting energy via the external field $\Delta(t)$, even though the total work is zero on average. The system of qubit and photons is thus an open system due to the presence of the dephasing environment and must be described by a mixed state in general. Nevertheless, this is not relevant when we focus on the scattered photons on the same frequency mode as the input. The averaged single-photon transmittance $\llangle t_\omega^{\mu} \rrangle$ and reflectance $\llangle r_\omega^\mu\rrangle$ are directly given by the pre-factors in Eq.~(\ref{AverageSfinal}) as} 
\begin{eqnarray}
\llangle t_\omega^\mu \rrangle = 1-\gamma_\mu{\cal L}[C_\phi(t)](\Gamma/2-i[\omega-\omega_0]),\label{AverageTransmittance}\\
\llangle r_\omega^\mu \rrangle = -\sqrt{\gamma_+\gamma_-}\llangle G_{\omega}\rrangle  = \sqrt{\gamma_{-\mu}/\gamma_\mu}\left(\llangle t_\omega^\mu \rrangle-1\right),\label{AverageReflectance}
\end{eqnarray}
\com{and measure the average amplitude of the photons} on the same $(\lambda=\mu)$ and opposite $(\lambda=-\mu)$ output channel. with respect to the input beam $\mu$. \com{Notice that the asymmetry in the couplings $\gamma_+\neq \gamma_-$ appears in Eqs.~(\ref{AverageTransmittance})-(\ref{AverageReflectance}) as a pre-factor of $\llangle G_{\omega}\rrangle$ and thus it only rescales the lineshape of the qubit. For the scope of the present paper it is therefore enough to consider examples in the symmetric case only ($\gamma_\mu=\gamma/2$), but we will still keep all the formulas general}.

\com{Equations (\ref{ScatteringEq})-(\ref{AverageReflectance})} have deep physical meaning as they allow us to predict the spectroscopic lineshape of a noisy qubit \com{either by solving the stochastic differential equation (\ref{ScatteringEq}) or by using} the knowledge of the Ramsey envelope $C_\phi(t)$ obtained \com{independently} via standard time-resolved noise experiments. \com{In Ref.~\cite{kubo63}, Kubo used the fluctuation-dissipation theorem to find a similar relation between $C_\phi(t)$ and the noise power spectrum, but this quantity is generic and not as simple to measure in a scattering experiment as the average transmittance we have introduced (see Sec.~\ref{sec:coherentextraction}).}  

Finally, we would like to remark that the present derivation may be \com{easily extended in various manners. So far we have considered a noisy qubit that is perfectly ``side-coupled'' to the waveguide, but in experiments there may be impedance mismatches and internal reflections that cause Fano resonance in the scattering profiles \cite{thyrrestrup17,javadi15}. Therefore, \ref{FanoRelations} generalizes Eqs.~(\ref{AverageTransmittance})-(\ref{AverageReflectance}) for a noisy qubit with a Fano resonance and shows that the corresponding relations between transmission and reflection coefficients are still valid under correlated dephasing noise. On the other hand, it is also possible to include multiple noise sources on the qubit. In this respect,} \ref{sec:WNbackground} shows that adding a white noise background $\Delta_{\rm WB}(t)$ to correlated noise, i.e.~$\Delta(t)\rightarrow\Delta(t)+\Delta_{\rm WB}(t)$, amounts to a trivial replacement $\Gamma/2\rightarrow \Gamma/2+\gamma_{\rm WB}$ in the stochastic equation (\ref{ScatteringEq}), where $\gamma_{\rm WB}$ is the pure dephasing rate of the white noise background. 

\subsection{Average transmittance of qubit with colored Gaussian dephasing}\label{LineShapeColoredGaussian}

In spectroscopy, correlated dephasing is typically referred to as \emph{spectral diffusion} \cite{berthelot06,coolen08,sallen10,wolters13} because it broadens the lineshapes of emitters. In this subsection we analyze this broadening and the average single-photon transmittance $\llangle t_\omega^{\mu} \rrangle$ of a qubit with colored Gaussian dephasing noise (see Sec.~\ref{IntrocoloredGaussian} for details on the model), paying special attention to the limits of white and quasi-static noise where the transmittance exhibits qualitatively different behaviors.

\begin{figure}[t!]
\center
\includegraphics[width=\linewidth]{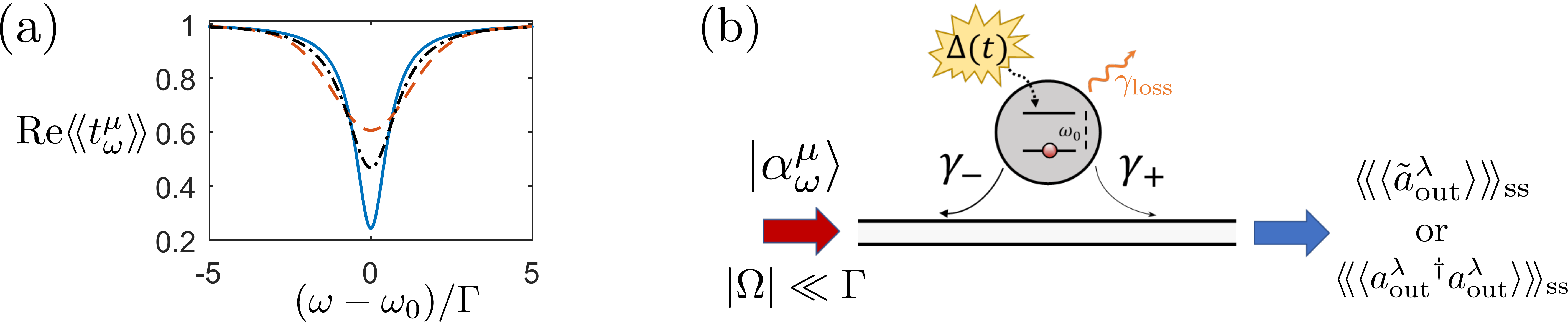}
\caption{(a) Average single-photon transmittance $\llangle t_\omega^{\mu} \rrangle$ of a noisy qubit coupled to a 1D photonic waveguide with parameters $\gamma_{\pm}=\gamma/2$, $\gamma=0.9\Gamma$, and $\gamma_{\rm loss}=0.1\Gamma$. We consider a colored Gaussian dephasing model, characterized by the correlation time $\tau_c=1/\kappa$ and the noise strength $\sigma=\Gamma$. For $\kappa=10\sigma$, the noise is in the white limit and the transmittance has a Lorentzian lineshape (blue/solid). For $\kappa=0$ the noise is quasi-static and $\llangle t_\omega^{\mu} \rrangle$ is Gaussian-like as given in Eq.~(\ref{QuasiStaticLineshape}) (red/dashed). Finally, for $\kappa=2\sigma$ the transmittance interpolates between two previous behaviors (black/dash-dotted). (b) Measurement of $\llangle t_\omega^{\mu} \rrangle$ of a noisy qubit in a waveguide, using a coherent state input \changed{($|\Omega|\ll \Gamma$)}, and homodyne or power measurements at the output.
\label{fig:spectroscopy}}
\end{figure}

Equation~(\ref{AverageTransmittance}) provides an expression for the single-photon transmittance $\llangle t_\omega^{\mu} \rrangle$ in terms of the analytical Ramsey envelope in Eq.~(\ref{OUexact}). For colored Gaussian noise with arbitrary correlation time $\tau_c=1/\kappa$ and noise strength $\sigma$ we can either estimate numerically the Laplace transform, or expand the super-exponential function in a power series to obtain 
\begin{eqnarray}
\llangle t_\omega^{\mu} \rrangle = 1-\gamma_\mu \sum_{n=0}^{\infty} \frac{(-1)^n}{n!} \frac{e^{(\sigma/\kappa)^2}(\sigma/\kappa)^{2n}}{\Gamma/2+\sigma^2/\kappa+n\kappa-i(\omega-\omega_0)}.\label{TransmittanceSeries}
\end{eqnarray}
In the limit of white noise \com{($\kappa,\sigma\rightarrow \infty$ with $\sigma^2/\kappa$ fixed)}, only the term with $n=0$ survives in Eq.~(\ref{TransmittanceSeries}), and the average transmittance is a Lorentzian function\footnote{In the white noise limit, Eq.~(\ref{OUexact}) becomes the exponential $C_\phi(t)={\rm exp}(-\gamma_\phi t)$, so that the Lorentzian lineshape follows directly from the Laplace transform in Eq.~(\ref{AverageTransmittance}).},
\begin{eqnarray}
\llangle t_\omega^{\mu} \rrangle=1-\frac{\gamma_\mu}{\Gamma/2+\gamma_\phi-i(\omega-\omega_0)},\label{LorentzianTransmittanceWN}
\end{eqnarray}
with pure dephasing rate $\gamma_\phi=\sigma^2/\kappa$. This is a well-known result, typically proven via the master equation formalism \cite{peropadre13}, which demonstrates that white noise \changed{pure} dephasing maintains the natural Lorentzian lineshape of the qubit, while its width and depth get modified by $\gamma_\phi$ \cite{astafiev10,wrigge08,thyrrestrup17}. This Lorentzian behavior is shown by the blue/solid transmittance in Figure \ref{fig:spectroscopy}(a), for typical waveguide QED parameters. If we now consider a finite but moderate correlation time $\sigma\lesssim\kappa<\infty$, more and more terms in the series expansion (\ref{TransmittanceSeries}) become important, resulting in a transmittance with a larger width and smaller depth, as shown by the black/dash-dotted curve in Figure \ref{fig:spectroscopy}(a). Finally, in the quasi-static limit of very long correlation times $\kappa\ll \sigma<\infty$, all terms in Eq.~(\ref{TransmittanceSeries}) contribute and the series expansion fails to converge numerically. In this case, we make the approximation $\kappa\rightarrow 0$ in which the relaxation function (\ref{OUexact}) is Gaussian, $C_\phi(t)={\rm exp}(-\sigma^2t^2/2)$, and perform the required Laplace transform analytically, obtaining\footnote{This expression can also be obtained by a simple static average over noiseless Lorentzian transmittances with different qubit frequencies as $\llangle t_\omega^\mu \rrangle =1-\gamma_\mu\int d\Delta {P}_{\rm G}(\Delta)/(\Gamma/2-i[\omega-\omega_0]+i\Delta)$, where ${P}_{\rm G}(\Delta)=(2\pi\sigma^2)^{-1/2}e^{-\Delta^2/(2\sigma^2)}$ is a Gaussian probability distribution with standard deviation $\sigma$.}
\begin{eqnarray}
\llangle t_\omega^{\mu} \rrangle & =1-\frac{\gamma_\mu}{\sigma}\sqrt{\frac{\pi}{2}}e^{\frac{\left(\Gamma/2-i[\omega-\omega_0]\right)^2}{2\sigma^2}}{\rm erfc}\left(\frac{\Gamma/2-i[\omega-\omega_0]}{\sqrt{2}\sigma}\right),\label{QuasiStaticLineshape}
\end{eqnarray}
with the \emph{complementary error function} ${\rm erfc}(z)=(2/\sqrt{\pi})\int_z^{\infty}dx e^{-x^2}$. From equation~(\ref{QuasiStaticLineshape}) we conclude that  in the slow noise limit $\kappa\ll \sigma<\infty$, $\llangle t_\omega^{\mu} \rrangle$ is Gaussian-like and has a width proportional to the noise strength $\sigma$. This behavior is shown by the red/dashed transmittance from figure \ref{fig:spectroscopy}(a). Notice that the Lorentzian (blue/solid) and Gaussian-like (red/dashed) lineshape limits can be qualitatively distinguished in transmittance experiments by their width, curvature, and tails \cite{petrakis67}, suggesting that spectroscopy can be a simple approach to discover the noise correlation properties. More specifically, fitting arbitrary parameters $\kappa$ and $\sigma$ to experimental transmittance data $\llangle t_\omega^{\mu} \rrangle$ one may even quantify the correlation time $\tau_c=1/\kappa$ and the noise strength $\sigma$ of a given environment as recently done in Ref.~\cite{eder18}.

Colored Gaussian noise is a useful and powerful dephasing model, and thus it is tempting to assume that this is the real noise. Indeed, this is what is done in most common waveguide-QED experiments, where a Lorentzian profile is assumed and a single dephasing parameter $\gamma_\phi$ is fitted\ \cite{astafiev10}. In Sec.~\ref{sec:coherentextraction} we will show that there is a more general approach, using estimates of the transmittance $\llangle t_\omega^{\mu} \rrangle$ to extract the Ramsey profile and noise correlations, in a single-photon scattering protocol that generalizes current experiments [see Figure \ref{fig:spectroscopy}(b)].

\subsection{Average transmittance of a qubit with 1/f dephasing noise}\label{LineShape1f}

In this subsection, \changed{we} consider a noisy qubit with dephasing due to \emph{1/f noise}, a very slowly varying, highly correlated, and low-frequency noise that is ubiquitously encountered in electronics and solid-state devices such as superconducting qubits or quantum dots \cite{Revpaladino14}. Nowadays there is still ongoing research on the microscopic origin and universal mechanisms behind this type of noise \cite{galperin06,koch07,eliazar10,quintana17,pachon17}, but an unquestionable experimental evidence is that its \emph{noise power spectrum}, ${\cal S}(\omega)=\sqrt{2\pi}{\cal F}\left[\llangle \Delta(0)\Delta(\tau)\rrangle\right](\omega)$, presents a power-law behavior ${\cal S}(\omega)\propto 1/\omega^\eta$, with $0<\eta<2$. In fact, it is exactly this low frequency divergence what makes 1/f noise so difficult to filter and to controllably observe in experiments \cite{Revpaladino14}.

\begin{figure}[t!]
\center
\includegraphics[width=\linewidth]{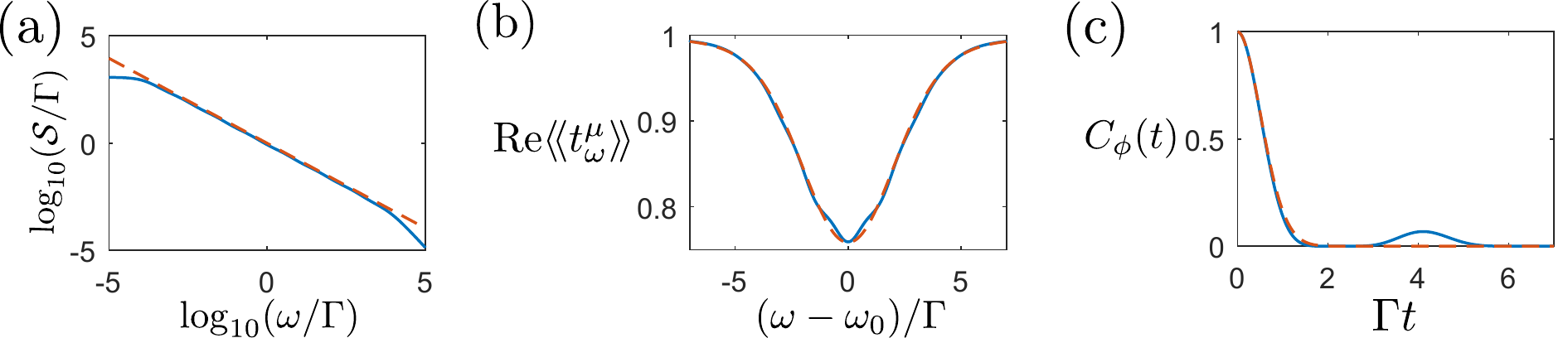}
\caption{Noisy qubit with 1/f dephasing noise. (a) Noise power spectrum ${\cal S}(\omega)$ for $1/f^{0.99}$ noise (red/dashed), and its simulation by $N=8$ uncorrelated noises (blue/solid) with $\kappa_1=10^{-5}\Gamma$, $\kappa_N=10\Gamma$, and $\sigma_1=2\Gamma$, as indicated in Eqs.~(\ref{idealS1f}) and (\ref{1fSpectrumExact}), respectively. (b) Average transmittance $\llangle t_\omega^\mu\rrangle$ for qubit with the $1/f^{0.99}$ dephasing noise model in (a), and the waveguide parameters $\gamma_{\pm}=\gamma/2$, $\gamma=0.9\Gamma$, and $\gamma_{\rm loss}=0.1\Gamma$. The red/dashed curve corresponds to Gaussian $1/f^{0.99}$ noise and the blue/solid to non-Gaussian $1/f^{0.99}$ noise caused $N=8$ different two-level fluctuators (see~\ref{sec:NoiseModels}). (c) Time-resolved Ramsey envelope $C_\phi(t)$ corresponding to the same parameters and same line-types as in (b).
\label{fig:Gaussian1f}}
\end{figure}

There have been various proposals for phenomenologically modeling the effects of 1/f noise within a finite but broad frequency window $\kappa_{\rm min}\ll \omega\ll \kappa_{\max}$ \cite{paladino02,shnirman05,galperin07,ruseckas10,kaulakys2005}. The basic assumption is that it is produced by a sum of $N$ uncorrelated noise sources,
\begin{eqnarray}
\Delta(t)=\frac{1}{\sqrt{N}}\sum_{j=1}^N \Delta_j(t),\label{uncorrelatedSum}
\end{eqnarray}
with noise components $\Delta_j(t)$ presenting correlations of the form $\llangle \Delta_j(0) \Delta_j(\tau)\rrangle=\sigma_j^2e^{-\kappa_j|\tau|}$, and thus the total autocorrelation and noise spectrum read 
\begin{eqnarray}
\fl\qquad\qquad\llangle \Delta(0)\Delta(\tau)\rrangle=\frac{1}{N}\sum_{j=1}^N\sigma_j^2 e^{-\kappa_j|\tau|},\qquad {\rm and}\qquad {\cal S}(\omega)=\frac{1}{N}\sum_{j=1}^N\frac{2\kappa_j\sigma_j^2}{\kappa_j^2+\omega^2}.\label{1fSpectrumExact}
\end{eqnarray}
To achieve this situation, the noise components $\Delta_j(t)$ can be modeled as independent Ornstein-Uhlenbeck processes \cite{kaulakys2005} (Sec.~\ref{IntrocoloredGaussian}), but it is also typically assumed that $\Delta_j(t)$ are originated by an ensemble of two-level fluctuators \cite{shnirman05,paladino02,galperin07}, characterized by different noise strengths $\sigma_j$ and jumping rates $\kappa_j$ (see \ref{sec:telegraph}). In either case, if the parameters $\kappa_j$ present an uniform distribution of $\log_{10}(\kappa_j/\Gamma)$ in a broad range from $\kappa_1=\kappa_{\rm min}$ to $\kappa_N=\kappa_{\rm max}$, and if $\sigma_j=\sigma_1\left(\kappa_1/\kappa_j\right)^{(\eta-1)/2}$, then in the limit $N\gg 1$ the noise spectrum ${\cal S}(\omega)$ in Eq.~(\ref{1fSpectrumExact})
approximates a power-law behavior \cite{kaulakys2005},
\begin{eqnarray}
{\cal S}(\omega)\approx \left[\frac{\pi \sigma_1^2 \kappa_1^{\eta-1}}{\sin(\pi\eta/2)\ln(\kappa_N/\kappa_1)}\right]\frac{1}{\omega^{\eta}},\qquad {\rm for}\quad \kappa_1\ll\omega\ll\kappa_N.\label{idealS1f}
\end{eqnarray}
In Figure~\ref{fig:Gaussian1f}(a) we illustrate the effectiveness of this method with a numerical simulation of $1/f^{0.99}$ noise with only $N=8$ independent noise components. We see that that exact noise spectrum in Eq.~(\ref{1fSpectrumExact}) (blue/solid), approximates well the expected the power-law behavior (red/dashed) in the frequency range $10^{-4}\ll\omega/\Gamma\ll 10^4$. 

Now we solve for the average transmittance $\llangle t_\omega^\mu\rrangle$ and the Ramsey envelope $C_\phi(t)$ for a noisy qubit subject to the above model of 1/f noise. We can simulate Gaussian or non-Gaussian 1/f noise depending if we choose the noise components $\Delta_j(t)$ in Eq.~(\ref{uncorrelatedSum}) as colored Gaussian noises (see Sec.~\ref{LineShapeColoredGaussian}) or as an ensemble of two-level fluctuators (see \ref{sec:1fnoise}). In the former case, $C_\phi(t)$ in Eq.~(\ref{gaussianCphi}) can be analytically computed \changed{from} the autocorrelation (\ref{1fSpectrumExact}), and reads
\begin{eqnarray}
C_{\phi}(t)={\rm exp}\left(-\frac{1}{N}\sum_{j=1}^N(\sigma_j/\kappa_j)^2(e^{-\kappa_j t}+\kappa_j t -1)\right),\label{Gaussian1fnoise}
\end{eqnarray} 
where $\kappa_j$ and $\sigma_j$ are chosen to simulate the 1/f model as explained above. To obtain $\llangle t_\omega^\mu\rrangle$ we use Eq.~(\ref{AverageTransmittance}) and numerically calculate the Laplace transform of Eq.~(\ref{Gaussian1fnoise}) as done in Sec.~\ref{IntrocoloredGaussian} for a single colored Gaussian noise. On the other hand, calculating $\llangle t_\omega^\mu\rrangle$ for non-Gaussian 1/f noise requires more advanced stochastic methods for describing the dynamics of the two-level fluctuators. This is done in full detail in \ref{sec:NoiseModels}, but here we discuss the results. In Figures~\ref{fig:Gaussian1f}(b) and (c) we display $\llangle t_\omega^\mu\rrangle$ and $C_\phi(t)$ for a noisy qubit with dephasing due to the $1/f^{0.99}$ noise simulated in \changed{Figure~\ref{fig:Gaussian1f}(a)}, and typical waveguide QED parameters. The red/dashed lines are the predictions in the Gaussian case as calculated via Eq.~(\ref{Gaussian1fnoise}), whereas the blue/solid lines correspond to the non-Gaussian situation, calculated from Eqs.~(\ref{VectorRealizations})-(\ref{VectorProb}). The main difference between Gaussian and non-Gaussian 1/f noises are small bumps in $\llangle t_\omega^\mu\rrangle$ and $C_\phi(t)$, which are signatures of the sparsity or granularity of the dephasing environment as treated in detail in \ref{sec:tunableGaussian}. Besides that, both predictions agree well and behave very similar to a single colored Gaussian noise in the quasi-static limit, except for the power-law spectrum ${\cal S}(\omega)$.

\section{Spectroscopic characterization of correlated dephasing noise}\label{sec:coherentextraction}

This section introduces a simple experimental protocol to measure the average single-photon transmittance and reflectance, and to recover the correlated dephasing noise from those quantities. This protocol only requires attenuated coherent states and either homodyne or power measurements at the output ---the choice of which depends mainly on whether the experiment is performed with microwave \cite{astafiev10,eder18} or optical photons \cite{javadi15,maser16,sipahigil16}---. \com{In Sec.~\ref{resultsProtocol} we summarize and discuss the most important results to apply the protocol, while Sec.~\ref{derivationProtocol} contains details on the derivation. In addition to this, \ref{FanoRelations} generalizes the protocol to the case the noisy qubit sees Fano resonances \cite{fano61,zhou08}, as for instance, in experiments with quantum dots in photonic crystals waveguides \cite{thyrrestrup17,javadi15,auffeves07}.}

\subsection{\com{Results of the protocol}}\label{resultsProtocol}

The experimental procedure is sketched in Figure \ref{fig:spectroscopy}(b), where a monochromatic coherent state $\ket{\alpha_\omega^\mu}$ of amplitude $\alpha_\omega^\mu$ and frequency $\omega$ is injected on the input channel $\mu=\pm$ of the waveguide. We study the evolution of the corresponding initial state
\begin{eqnarray}
\ket{\Psi(0)}=\ket{\Psi_{\rm qb}}\ket{\alpha_\omega^\mu},\label{iniState}
\end{eqnarray}
which describes a coherently driven qubit from an arbitrary initial state $\ket{\Psi_{\rm qb}}$, and vacuum states in all photonic channels different than $\mu$. We will work in the limit of weak driving \changed{$|\Omega|\ll \Gamma$,} with driving strength given by $\Omega = -i\alpha_\omega^{\mu} \sqrt{\gamma_\mu}$. We show below that in this limit we recover the single-photon transmittance from homodyne or power measurements \changed{in steady state}, as follows
\begin{eqnarray}
\frac{\llangle \braket{\tilde{a}^{\mu}_{\rm out}} \rrangle_{\rm ss}}{\alpha_\omega^\mu}= \llangle t_\omega^\mu \rrangle + {\cal O}\left[|\Omega|/\Gamma \right]^2,\label{TMeasurementHomodyne}\\
\frac{\llangle \braket{a^{\mu}_{\rm out}{}^\dag a^{\mu}_{\rm out}} \rrangle_{\rm ss}}{|\alpha_\omega^\mu|^2} = 2\beta_\mu-1+2\left(1-\beta_\mu\right){\rm Re}\lbrace \llangle t_\omega^{\mu}\rrangle\rbrace+{\cal O}\left[|\Omega|/\Gamma\right]^2,\label{TransmittanceMeasurement}
\end{eqnarray}
with $\beta_\mu=\gamma_\mu/\Gamma$ the directional $\beta$-factor, \com{and $\tilde{a}_{\rm out}^\mu(t) = e^{i\omega t} a_{\rm out}^\mu(t)$.}

Note that, while homodyne measurements \changed{in Eq.~(\ref{TMeasurementHomodyne})} provide direct access to $\llangle t_\omega^{\mu}\rrangle$, power measurements \changed{in Eq.~(\ref{TransmittanceMeasurement})} give us only its real part, but we can still reconstruct the full transmittance via the Kramers-Kronig relation,
\begin{eqnarray}
{\rm Im}\lbrace \llangle t_\omega^\mu \rrangle \rbrace = \frac{1}{\pi}{\cal P}\int_{-\infty}^\infty d\omega' \frac{1-{\rm Re}\lbrace \llangle t_{\omega'}^\mu \rrangle \rbrace}{\omega'-\omega},\label{KK}
\end{eqnarray}
with ${\cal P}$ representing the Cauchy's principal value of the integral. \com{In the literature it is not well recognized that power measurements alone are enough to determine the modulus and phase of the transmittance $\llangle t_\omega^\mu \rrangle$, even in the presence of general correlated noise and dissipation as shown here. Indeed, it is typically believed that power measurements give direct access to $|\llangle t_\omega^\mu \rrangle|^2$, but in \ref{ReflectionExpressions} we show that this is only true in the absence of any dephasing, so that ${\rm Re}\lbrace \llangle t_\omega^\mu \rrangle\rbrace=|\llangle t_\omega^\mu \rrangle|^2$. For more details see \ref{ReflectionExpressions}, which also includes the expressions for single-photon reflectance measurements $\llangle r_\omega^\mu \rrangle$, and a discussion on the conservation of the average photon flux in these experiments.}  

\com{After measuring the single-photon transmittance $\llangle t_\omega^\mu \rrangle$,} we can invert equation~(\ref{AverageTransmittance}) to access to the time-resolved Ramsey envelope $C_\phi(t)$ and characterize noise correlations of the environment. A convenient inverse formula can be derived when the dephasing fluctuation $\Delta(t)$ has a symmetric probability distribution around the average, which is very reasonable assumption in experiments. In this case, the Ramsey envelope defined in Eq.~(\ref{DephasingRelaxation}) is a real function of time, $C_\phi(t)=[C_\phi(t)]^\ast$, and it can be directly related to ${\rm Re}\llangle t_\omega^\mu \rrangle$ by\footnote{Inverting Eq.~(\ref{AverageTransmittance}) in the general case leads to $C_\phi(t)=(2\pi)^{-1/2}e^{(\Gamma/2)t}{\cal F}^{-1}\left[(1-\llangle t_\omega^\mu\rrangle)/\gamma_\mu\right](t)$, for $t>0$, but this expression presents a slower numerical convergence compared to Eq.~(\ref{realRelation}). Notice that the presence of a non-zero photon decay $\Gamma>0$ allows us to mathematically replace the inverse Laplace transform ${\cal L}^{-1}$ by the more convenient inverse Fourier transform ${\cal F}^{-1}$.}
\begin{eqnarray}
C_\phi(t)=\sqrt{\frac{2}{\pi}}e^{(\Gamma/2)t}{\cal F}^{-1}\left[\frac{1-{\rm Re}\llangle t_\omega^\mu \rrangle}{\gamma_\mu}\right](t),\quad {\rm for}\quad t>0.\label{realRelation}
\end{eqnarray}
This relation (\ref{realRelation}) has important physical consequences to single-photon scattering experiments in waveguide QED, as it demonstrates that applying a Fourier transformation on the usual transmittance data \cite{astafiev10,wrigge08,thyrrestrup17,eder18,javadi15,maser16,sipahigil16}, one can characterize noise correlations without requiring direct access and time-dependent control of the emitter. Moreover, Eq.~(\ref{realRelation}) is particularly convenient in the case of power measurements (\ref{TransmittanceMeasurement}) as it only requires the knowledge of ${\rm Re}\llangle t_\omega^\mu \rrangle$, and thus avoids the use of the Kramers-Kronig transformation (\ref{KK}).

\subsection{\com{Derivation of the protocol}}\label{derivationProtocol}

Let us briefly summarize how equations  (\ref{TMeasurementHomodyne})-(\ref{TransmittanceMeasurement}) are derived. We begin with the equations of motion for the noisy qubit, taking expectation values on (\ref{EqMotSm})-(\ref{EqMotSz}) with the initial condition (\ref{iniState}). Using the property $a_{\rm in}^\lambda(t)\ket{\Psi(0)}=\alpha_\omega^\mu \delta_{\lambda\mu}e^{-i\omega t}\ket{\Psi(0)}$ with $\lambda=\pm$, and going to a rotating frame with the driving frequency $\omega$, we find
\begin{eqnarray}
\frac{d}{dt}\braket{\tilde{\sigma}^{-}}&=&-\left(\frac{\Gamma}{2}-i[\omega-\omega_0]+i\Delta(t)\right)\!\braket{\tilde{\sigma}^{-}}-\Omega\braket{\sigma_z},\label{BlochSigmamin}\\
\frac{d}{dt}\braket{\sigma_z}&=&-\Gamma(1+\braket{\sigma_z})+2(\Omega^\ast \braket{\tilde{\sigma}^{-}}+{\rm h.c.}).\label{BlochSigmaZ}
\end{eqnarray}
Here, we have defined the slowly evolving coherence $\langle\tilde{\sigma}^{-}(t)\rangle=e^{i\omega t}\langle\sigma^{-}(t)\rangle$ and the strength of the coherent drive $\Omega = -i\alpha_\omega^{\mu} \sqrt{\gamma_\mu}$. The qubit equations are stochastic Bloch equations that can combine correlated dephasing with saturation at strong drives $|\Omega|\gtrsim \Gamma$ \cite{fan2012}. The stochastic methods from section~\ref{sec:NoiseModels} provide a solution to this complex dynamics of the qubit, but noise spectroscopy only requires the steady state averaged coherence $\llangle \braket{\tilde{\sigma}^-}\rrangle_{\rm ss}=\llangle \braket{\tilde{\sigma}^-}\rrangle_{t\rightarrow\infty}$, which appears both in homodyne and power steady state measurements as\footnote{To derive the relation (\ref{power1}), we combined the input-output equations (\ref{InputOutput}), the equation of motion (\ref{BlochSigmaZ}), and used the exact relation $\llangle\braket{\sigma_z}\rrangle_{\rm ss}=-1+{\rm Re}\left\lbrace 4 \Omega^\ast \llangle\braket{\tilde{\sigma}^-}\rrangle_{\rm ss}/\Gamma \right\rbrace$, which results from integrating~(\ref{BlochSigmaZ}) and averaging in steady state.}
\begin{eqnarray}
&\frac{\llangle \braket{\tilde{a}_{\rm out}^\lambda}\rrangle_{\rm ss}}{\alpha_\omega^\mu}=\delta_{\lambda\mu} - \sqrt{\gamma_\lambda\gamma_\mu} \llangle\braket{\tilde{\sigma}^-}\rrangle_{\rm ss}/\Omega,\label{homodyne1}\\
&\frac{\llangle\braket{a_{\rm out}^\lambda{}^\dag a_{\rm out}^\lambda}\rrangle_{\rm ss}}{|\alpha_\omega^\mu|^2} = \delta_{\lambda\mu} -2\sqrt{\gamma_\lambda\gamma_\mu}\left(\delta_{\lambda\mu} -\sqrt{\beta_\lambda\beta_\mu}\right){\rm Re}\left\lbrace \llangle\braket{\tilde{\sigma}^-}\rrangle_{\rm ss}/\Omega\right\rbrace.\label{power1}
\end{eqnarray}
In the low driving limit $|\Omega|\ll \Gamma$, the qubit will remain close to the ground state $\braket{\sigma_z}=-1+{\cal O}\left[|\Omega|/\Gamma\right]^2$, and equations~(\ref{BlochSigmamin}) and (\ref{ScatteringEq}) become equivalent. We can thus map the qubit steady state coherence $\llangle\braket{\tilde{\sigma}^-}\rrangle_{\rm ss}$ to the solution of average scattering overlap in Sec.~\ref{sec:scattering} as,
\begin{eqnarray}
\llangle\braket{\tilde{\sigma}^-}\rrangle_{\rm ss}/\Omega = \llangle G_\omega \rrangle + {\cal O}\left[|\Omega|/\Gamma\right]^2. \label{Sconnection}
\end{eqnarray}
From this relation we conclude that homodyne and power measurements give us full information about the average single-photon transmittance $\llangle t_\omega^\mu \rrangle$ and reflectance $\llangle r_\omega^{\mu}\rrangle$, and allow a full spectroscopic characterization of the noise via Eqs.~(\ref{TMeasurementHomodyne})-(\ref{realRelation}).

\section{Conclusions and outlook}\label{sec:summary}

We \com{developed a stochastic version of input-output theory which consistently describes} the effect of correlated dephasing noise in single-photon scattering experiments with weak coherent inputs. Using this theory, we studied scattering subject to the typical noise models from solid-state and quantum optics ---white noise, quasi-static noise, colored Gaussian noise (see Secs.~\ref{IntrocoloredGaussian} and \ref{LineShapeColoredGaussian}), and 1/f noise (see Sec.~\ref{LineShape1f}), in addition to telegraph noise and non-Gaussian jump models in \ref{sec:NoiseModels}---, illustrating how to calculate the single-photon transmittance $\llangle t_\omega^\mu\rrangle$ and reflectance $\llangle r_\omega^{\mu}\rrangle$ of each model.

Complementary to these theoretical developments, we introduced a spectroscopic method that extracts the qubit noise correlations from standard homodyne or photon counting measurements. The method provides the same information as time-resolved Ramsey experiments, but does not require direct access or time-dependent control of the emitter. This method and the techniques developed in this work are suited not only for waveguide QED experiments ---superconducting circuits \cite{astafiev10,hoi13,forn-diaz17,eder18}, quantum dots in photonic crystals \cite{thyrrestrup17,javadi15}, SiV-centers in diamond waveguides \cite{sipahigil16,lemonde18}, or nanoplasmonics \cite{martincano14}---, but also for generic experiments with two-level quantum emitters interacting with propagating photons, such as molecules in a 3D bath \cite{maser16} or ions in a Paul trap \cite{araneda18}.

There are still several open questions and extensions to consider in the interaction between few photons and noisy quantum emitters. For instance, our theory is valid for general stationary random fluctuations $\Delta(t)$, but we only analyzed \changed{phenomenological} classical noise models. Therefore, it would be interesting to study the effects of specific microscopic quantum models producing correlated \changed{pure} dephasing on the quantum scatterers \cite{suter16,nemet18}, and try to find the connections \changed{to the phenomenological models analyzed here}. Moreover, we can combine the stochastic methods discussed here with our recent theory of scattering tomography \cite{ramos17} to characterize multi-photon processes or many-body scatterers \cite{pazsilva17,szankowski16,prasanna18} under realistic conditions of noise.

\ack

\com{The authors acknowledge discussions with P. Eder and F. Deppe. This work was supported by} the MINECO/FEDER Project FIS2015-70856-P and CAM PRICYT Research Network QUITEMAD+ S2013/ICE-2801. TR further acknowledges the Juan de la Cierva fellowship FJCI-2016-29190.

\appendix

\section{Average single-photon transmittance of a qubit with dephasing due to correlated non-Gaussian Markovian noise models}\label{sec:NoiseModels}

In the main text we explicitly calculated the average transmittance of a qubit with colored Gaussian noise and Gaussian 1/f noise, where $C_\phi(t)$ is analytical and $\llangle t_\omega^\mu \rrangle$ can be directly obtained from (\ref{AverageTransmittance}). Although Gaussian noise models are very successfully applied in numerous experiments \cite{suter16}, the Gaussianity assumption breaks down in situations where the qubit is coupled to a sparse dephasing environment \cite{norris16} such as a few frequency modes \cite{kotler13}, or ensembles of few \changed{two-level fluctuators (TLFs)} \cite{galperin06,galperin07}. In the following we extend the analysis to arbitrary correlated non-Gaussian Markovian noise models which include telegraph noise caused by a single TLF (see \ref{sec:telegraph}), tunable non-Gaussian noise caused by a sparse ensemble of TLFs (see \ref{sec:tunableGaussian}), and non-Gaussian 1/f noise (see \ref{sec:1fnoise}) typically found in solid-state devices \cite{Revpaladino14}. To compute $\llangle t_\omega^\mu \rrangle$ and $C_\phi(t)$ for a qubit under these types of dephasing, we require the stochastic methods introduced in the following subsection \ref{SolvingMain}.

\subsection{Stochastic differential equations with arbitrary correlated Markovian noise}\label{SolvingMain}

Here we state the equations to solve for the average transmittance $\llangle t_\omega^\mu\rrangle$ and the Ramsey envelope $C_\phi(t)$ in the case of the most general correlated, stationary, and Markovian dephasing noise. In practice, we \changed{generalize} the method in page 418 of Ref.~\cite{vankampen92} to inhomogeneous stochastic differential equations, and then apply it to the scattering equation (\ref{ScatteringEq}).

Our first assumption is that the stochastic process $\Delta(t)$ is stationary and Markovian. The probability for the noise to be in realization $\Delta(t)=\Delta$ at time $t$, conditioned on being $\Delta(t_0)=\Delta_0$ at time $t_0$ is denoted by $P(\Delta,t)=P(\Delta,t|\Delta_0,t_0)$. The most general Markovian dynamics for the above conditional probability is governed by a differential Chapman-Kolmogorov equation \cite{gardiner1985},
\begin{eqnarray}
\frac{\partial}{\partial t}P(\Delta,t)= LP(\Delta,t),\label{generalLiovillian}
\end{eqnarray}
with initial condition $P(\Delta,t_0)=\delta(\Delta-\Delta_0)$ and classical Liouvillian $L$ given by
\begin{eqnarray}
\fl\quad LP(\Delta,t) = -\frac{\partial}{\partial \Delta}[D_1(\Delta) P(\Delta,t)]+\frac{1}{2}\frac{\partial^2}{\partial \Delta^2}[D_2(\Delta) P(\Delta,t)]+\int d\Delta' W(\Delta,\Delta')P(\Delta',t).\label{ChapmanKolmogorov}
\end{eqnarray}
Here, $D_1(\Delta)$ is the drift function, $D_2(t)\geq 0$ the diffusion function, and $W(\Delta,\Delta')\geq 0$ for $\Delta\neq\Delta'$ are transition probabilities between different values of the noise. The conservation of total probability also requires $\int d\Delta\ L P(\Delta,t)=0$ and thus $\int d\Delta\ W(\Delta,\Delta')=0$. We further assume $L$ is time-independent to have an homogeneous stationary Markovian process with well-defined steady state $LP_{\rm ss}(\Delta)=0$.

We want to study $G_\omega(t)$ which is a stochastic process related to $\Delta(t)$ via Eq.~(\ref{ScatteringEq}). Since $\Delta(t)$ is Markovian, the joint process $[\Delta(t),G_\omega(t)]$ is Markovian too \cite{vankampen92} with joint probability denoted by ${\cal P}(G_\omega,\Delta,t)$. For a multiplicative inhomogeneous stochastic differential equation of the form $dG/dt=A(\Delta)G+B$, the joint probability satisfies \cite{vankampen92}
\begin{eqnarray}
\frac{\partial}{\partial t}{\cal P}(G_\omega,\Delta,t)= -A(\Delta)\frac{\partial}{\partial G_\omega}(G_\omega {\cal P})-B\frac{\partial {\cal P}}{\partial G_\omega} + L{\cal P},\label{FokkerPlanckJointApp}
\end{eqnarray}
with the initial condition ${\cal P}(G_\omega,\Delta,0)=\delta(G_\omega-G_\omega(0))P(\Delta,0)$. To compute the noise average $\llangle G_\omega\rrangle$, the strategy is to convert the stochastic equation (\ref{ScatteringEq}) into a set of ordinary differential equations for the \emph{marginal averages} $g_{\omega}(\Delta,t)=\int dG_\omega G_\omega {\cal P}(G_\omega,\Delta,t)$, and from its solution obtain the total average as $\llangle G_\omega \rrangle (t)=\int d\Delta g_{\omega}(\Delta,t)$. To do so, we \changed{insert} $A(\Delta)=-[\Gamma/2-i(\omega-\omega_0)+i\Delta]$ and $B=1$ in Eq.~(\ref{FokkerPlanckJointApp}), multiply it by $G_\omega$ and integrate it over $G_\omega(t)$, obtaining
\begin{eqnarray}
\fl\qquad\quad\frac{\partial g_{\omega}(\Delta,t)}{\partial t}=&-[\Gamma/2-i(\omega-\omega_0)+i\Delta]g_{\omega}(\Delta,t)+P(\Delta,t)+Lg_\omega(\Delta,t).\label{MarginalGeneral}
\end{eqnarray}
For the scattering problem in Sec.~\ref{averageSMatrix}, the differential equation (\ref{MarginalGeneral}) must be solved with the initial condition $g_{\omega}(\Delta,-\infty)=G_\omega(-\infty)P(\Delta,-\infty)=0$, which effectively corresponds to finding the steady state solution $g_\omega^{\rm ss}(\Delta)=g_\omega(\Delta,t\rightarrow\infty)$ or $dg_{\omega}(\Delta,t)/dt=0$. Finally, when having the steady state marginal averages $g_\omega^{\rm ss}(\Delta)$ for each frequency $\omega$ and each noise realization $\Delta$, we obtain the average transmittance as
\begin{eqnarray}
\llangle t_\omega^\mu\rrangle = 1 - \gamma_\mu \int d\Delta g_\omega^{\rm ss}(\Delta).\label{transmittanceIntegration} 
\end{eqnarray}
\changed{We} see that computing the average transmittance $\llangle t_\omega^\mu\rrangle$ for the most general non-Gaussian, correlated, stationary, and Markovian noise model amounts to solve for the steady state of the partial differential equation (\ref{MarginalGeneral}) and then to integrate it in Eq.~(\ref{transmittanceIntegration}) over all noise realizations. 

\changed{To simplify the above} solution, we now particularize the analysis to discrete jump noise models, where the stochastic process $\Delta(t)$ has a discrete number of realizations denoted by $\Delta_m$. In this case, we can set $D_1=D_2=0$ in Eq.~(\ref{ChapmanKolmogorov}), and the probability $P(\Delta_m,t)$ for the noise to be in the realization $\Delta(t)=\Delta_m$ at time $t$, conditioned on being $\Delta(t_0)=\Delta_{m_0}$ at $t=t_0$ is governed by the time-local rate equation \cite{vankampen92},
\begin{eqnarray}
\frac{d}{dt}P(\Delta_m,t)=LP(\Delta_m,t)=\sum_{n}W_{mn}P(\Delta_n,t).\label{RateEqDiscrete}
\end{eqnarray}
Here, the matrix coefficients $W_{mn}\geq 0$ for $m\neq n$ describe transition rates of the noise to jump from realization $\Delta_n$ to $\Delta_m$ ---which must satisfy $\sum_{m} W_{mn}=0$ to ensure the conservation of total probability $\sum_m P(\Delta_m,t)=1$---. Importantly, the partial differential equation (\ref{MarginalGeneral}) reduces to a discrete set of ordinary differential equations for the discrete number of marginal averages $g_{\omega}(\Delta_m,t)$ as,
\begin{eqnarray}
\fl\ \ \frac{d}{dt}g_{\omega}(\Delta_m,t)=&-\left(\frac{\Gamma}{2}-i[\omega-\omega_0]+i\Delta_m\right)g_{\omega}(\Delta_m,t)+P(\Delta_m,t)+\sum_n W_{mn}g_{\omega}(\Delta_n,t),\label{MarginalsEqDiscrete}
\end{eqnarray}
which now allows us for a much simpler steady state solution. In fact, setting $dg_{\omega}(\Delta_m,t)/dt=0$ in Eq.~(\ref{MarginalsEqDiscrete}) we can map the problem to a linear system of equations,
\begin{eqnarray}
\sum_{n} J_{mn} g_\omega^{\rm ss}(\Delta_n)=P_{\rm ss}(\Delta_m),\qquad {\rm with}\label{LinearSystemEq}\\
J_{mn}=[\Gamma/2-i(\omega-\omega_0)+i\Delta_m]\delta_{mn}-W_{mn}.\label{MmatrixLS}
\end{eqnarray}
Here, the matrix $J_{mn}$ is of the same size as $W_{mn}$, and $P_{\rm ss}(\Delta_m)$ denotes the steady state solution of the rate equations (\ref{RateEqDiscrete}). Finally, solving this linear problem for different values of the input field $\omega$, we obtain the average single-photon transmittance from the sum,
\begin{eqnarray}
\llangle t_\omega^\mu\rrangle = 1 - \gamma_\mu \sum_m g_\omega^{\rm ss}(\Delta_m).\label{TransmittanceDiscrete}
\end{eqnarray}

On the other hand, to obtain the Ramsey envelope $C_\phi(t)$ in Eq.~(\ref{DephasingRelaxation}), we can numerically extract it from $\llangle t_\omega^\mu\rrangle$ via the inversion formula (\ref{realRelation}). Alternatively, we can also obtain it by calculating the average solution $C_\phi(t)=\llangle X (t)\rrangle$ of the homogeneous stochastic differential equation, 
\begin{eqnarray}
\frac{d}{dt}X(t)=-i\Delta(t)X(t).
\end{eqnarray}
A set of differential equations for the marginal averages \changed{$x(\Delta_m,t)=\int dX {\cal P}(X,\Delta_m,t)X$} can be derived from Eq.~(\ref{FokkerPlanckJointApp}) with $A(\Delta)=-i\Delta$, $B=0$, and the discrete rate equations (\ref{RateEqDiscrete}),
\begin{eqnarray}
\fl\qquad\quad\frac{d}{dt}x(\Delta_m,t) = -i\Delta_m x(\Delta_m,t)+\sum_{n}W_{mn}x(\Delta_n,t),\label{TimeDepRamsey}
\end{eqnarray}
which must be solved for the initial condition $x(\Delta_m,0)=P_{\rm ss}(\Delta_m)$. Finally, we obtain the Ramsey envelope as $C_\phi(t)=\llangle X (t)\rrangle=\sum_m x(\Delta_m,t)$. 

In the following three subsections, we evaluate $\llangle t_\omega^\mu\rrangle$ and $C_\phi(t)$ for different forms and sizes of $W_{mn}$ corresponding to correlated telegraph noise, and more general non-Gaussian 1/f noise models.

\subsection{Telegraph correlated noise}\label{sec:telegraph}

Charges or impurities in the materials of solid-state devices are modeled in many cases as localized double-well potentials or two-level fluctuators (TLFs) \cite{Revpaladino14,galperin06,phillips87,ramos13}. A strong resonant coupling between the qubit and an environmental TLS can lead to the observation of resonances \cite{lisenfeld16,simmonds04,astafiev04}, but a weak off-resonant coupling can induce fluctuating Stark shifts on the qubit and thus originate correlated dephasing as in Eq.~(\ref{Hqubit}). 
Although TLFs naturally appear in large ensembles of them \cite{shnirman05,paladino02,galperin07}, the \emph{telegraph noise} produced by a single TLS is an instructive and exactly solvable model capturing many features of more complex correlated non-Gaussian noises.

\begin{figure}[t!]
\center
\includegraphics[width=1\linewidth]{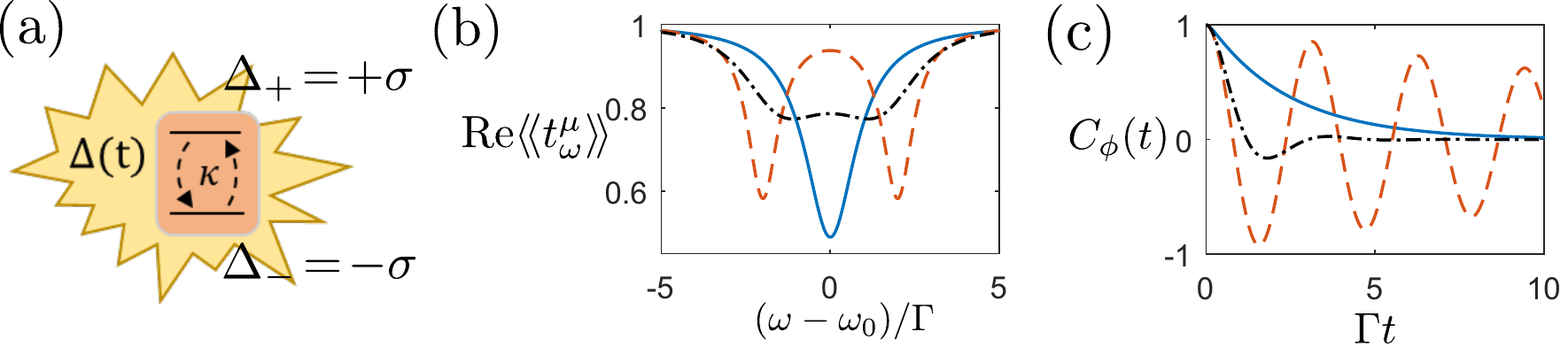}
\caption{Single-photon scattering on a noisy qubit with random telegraph dephasing. (a) Scheme of two-level fluctuator (TLF) randomly changing the qubit resonance as $\omega_0\pm\sigma$ with a rate $\kappa$. (b) Predictions for the average transmittance $\llangle t_\omega^\mu\rrangle$ for $\kappa=5\sigma$ (white noise, blue/solid), $\kappa=\sigma$ (black/dash-dotted), and $\kappa=0.05\sigma$ (quasi-static, red/dashed). Other parameters are $\sigma=2\Gamma$, $\gamma_{\pm}=\gamma/2$, $\gamma=0.9\Gamma$, and $\gamma_{\rm loss}=0.1\Gamma$. (c) Time-resolved Ramsey envelope $C_\phi(t)$ corresponding to the same parameters and same line-types as in (b).  
\label{fig:TelegraphNoise}}
\end{figure}

Telegraph noise is the simplest jump model, where random variable $\Delta(t)$ can only take two possible values $\Delta_\pm=\pm \sigma$ \cite{jacobs2010,gardiner1985}, corresponding to an increase or decrease of the qubit resonance as $\omega_0\pm\sigma$. The dynamics of this noise consists in random jumps with rate $\kappa$ between the two possible realizations $\Delta_m$ with $m=\pm$, as depicted in Figure \ref{fig:TelegraphNoise}(a). The probabilities $P(\Delta_m,t)$ of being in $\Delta_m$ at time $t$, conditioned of being in $\Delta_{m_0}$ at an initial time $t_0$, are governed by the Markovian rate equations,
\begin{eqnarray}
\frac{d}{dt} P(\Delta_m,t)=-\frac{\kappa}{2}P(\Delta_m,t)+\frac{\kappa}{2}P(\Delta_{-m},t),\label{MarkovianRateTelegraph}
\end{eqnarray}
which can be recast in the general form (\ref{RateEqDiscrete}) with the transition matrix $W_{mn}=-mn\kappa/2$ ($m,n=\pm$). The above equations imply that in steady state the probabilities of being in either realization are equal $P_{\rm ss}(\Delta_m)=1/2$, the mean fluctuation vanishes $\llangle \Delta(t)\rrangle=0$, and the autocorrelation has the same form $\llangle \Delta(0) \Delta(\tau)\rrangle=\sigma^2 e^{-\kappa |\tau|}$ \cite{jacobs2010,gardiner1985} as the colored Gaussian noise in Eq.~(\ref{OUcorrelation}). Notice that this is just a coincidence since higher order correlations highly differ due to the non-Gaussian character of the telegraph noise \cite{gardiner1985}.

The simplicity of the telegraph noise allows us to analytically solve for the average transmittance $\llangle t_\omega^\mu\rrangle$ in Eq.~(\ref{TransmittanceDiscrete}), since the linear system (\ref{LinearSystemEq}) is of size 2-by-2 with the matrix $J_{mn}=[\Gamma/2-i(\omega-\omega_0)+im\sigma]\delta_{mn}+mn\kappa/2$ ($m,n=\pm$), and $P_{\rm ss}(\Delta_m)=1/2$. For the steady state marginal averages we obtain
\begin{eqnarray}
g^{\rm ss}_\omega(\Delta_m)=\frac{(\Gamma/2-i[\omega-\omega_0]+\kappa-im\sigma)}{2\left[(\Gamma/2-i[\omega-\omega_0]+\kappa/2)^2+\sigma^2-\kappa^2/4\right]},
\end{eqnarray}
and using Eq.~(\ref{TransmittanceDiscrete}) we find that $\llangle t_\omega^\mu\rrangle$  can be expressed in a form reminiscent to a Lorentzian, $\llangle t_\omega^\mu\rrangle=1-\gamma_\mu/[\Gamma/2+\gamma_\phi(\omega)-i(\omega-\omega_0)]$, but with a frequency-dependent pure dephasing rate $\gamma_\phi(\omega)$ given by
\begin{eqnarray}
\gamma_\phi(\omega)=\frac{\sigma^2}{\Gamma/2+\kappa-i(\omega-\omega_0)}.\label{freqDepGammaPhi}
\end{eqnarray}
The lineshape is thus not Lorentzian in general, except for the white noise limit, \com{($\kappa,\sigma\rightarrow\infty$ with $\sigma^2/\kappa$ constant) where the dephasing rate (\ref{freqDepGammaPhi}) becomes the constant $\gamma_\phi(\omega)=\sigma^2/\kappa$.} This is illustrated by the blue/solid transmittance in Fig.~\ref{fig:TelegraphNoise}(b) for standard waveguide QED parameters. For a finite but moderate correlation time $\sigma\lesssim\kappa<\infty$, $\llangle t_\omega^\mu\rrangle$ gets broader than the Lorentzian (black/dash-dotted), and in the quasi-static limit of long correlation times $\kappa\ll \sigma<\infty$, $\llangle t_\omega^\mu\rrangle$ develops two well separated dips centered at $\omega\approx\omega_0\pm\sigma$ whose widths are proportional to $\sigma$ (red/dashed).

To obtain the Ramsey envelope $C_\phi(t)$ for the qubit under this telegraph noise, we can either use the inverse relation Eq.~(\ref{realRelation}) on our known $\llangle t_\omega^\mu\rrangle$ or solve the differential Eq.~(\ref{TimeDepRamsey}), which gives
\begin{eqnarray}
C_{\phi}(t)=\frac{1}{2}\left[(1+v_0)e^{v_{+}t}+(1-v_0)e^{v_{-}t}\right],
\end{eqnarray}
with $v_0=\kappa/\sqrt{\kappa^2-4\sigma^2}$ and $v_{\pm}=(-\kappa\pm \sqrt{\kappa^2-4\sigma^2})/2$ \cite{vankampen92}. As shown in Figure \ref{fig:TelegraphNoise}(c), $C_\phi(t)$ is the exponential decay in the white noise limit, and for a finite correlation time $\kappa<\infty$, it shows damped oscillations with frequency $\sim \sigma$ and damping rate $\sim \kappa$.

\subsection{Correlated dephasing noise with tunable non-Guassianity}\label{sec:tunableGaussian}

In this subsection we introduce a model of non-Gaussian correlated noise, whose non-Gaussianity can be tuned to describe situations such as the telegraph noise from previous subsection, all the way to the limit of colored Gaussian noise in Secs.~\ref{IntrocoloredGaussian} and \ref{LineShapeColoredGaussian}. 

\begin{figure}[t!]
\center
\includegraphics[width=1\linewidth]{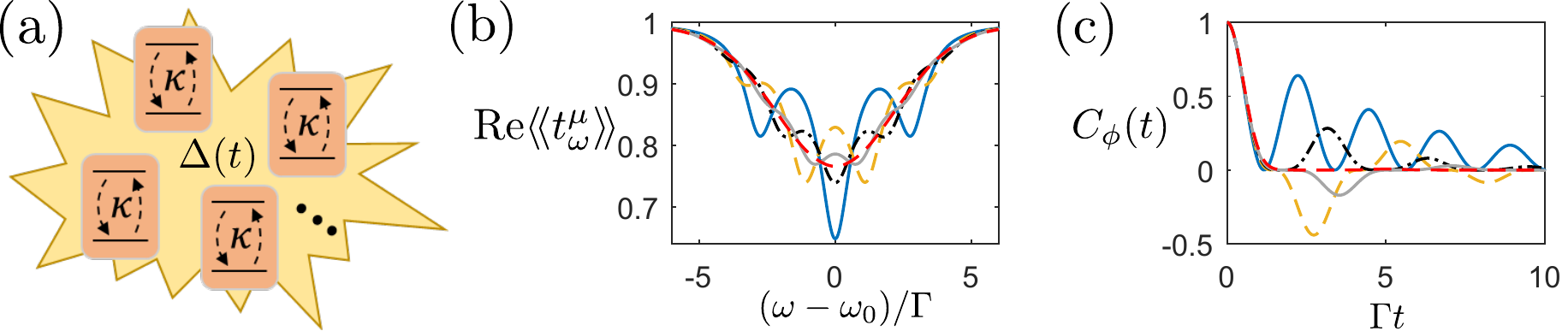}
\caption{Noisy qubit with non-Gaussian noise due to an ensemble of $M$ identical and independent two-level fluctuators (TLFs). (a) Scheme of the dephasing environment, characterized by jumps at rate $\kappa$ and an average noise amplitude $\sigma$. (b) Average transmittance $\llangle t_\omega^\mu\rrangle$ for $M=[2,3,4,5,10]$ (blue, orange, black, grey, red), and the parameters $\kappa=0.1\sigma$, $\sigma=2\Gamma$, $\gamma_{\pm}=\gamma/2$, $\gamma=0.9\Gamma$, and $\gamma_{\rm loss}=0.1\Gamma$. (c) Time-resolved Ramsey envelope $C_\phi(t)$ corresponding to the same parameters and same line-types as in (b).\label{fig:ResultsnonGaussianNoise}}
\end{figure}

We follow Ref.~\cite{hu2015} and construct a discrete noise model from the sum of $M$ independent and identical TLFs, $\Delta(t)=\sum_{l=1}^M \Delta_l(t)/\sqrt{M}$ (see Figure~\ref{fig:ResultsnonGaussianNoise}(a)). Here, each noise component $\Delta_l(t)$ corresponds to a telegraph noise as in the previous subsection, which flips between the values $\Delta_l(t)=\pm\sigma$ at a rate $\kappa$ and independently satisfies the Markovian rate equation (\ref{MarkovianRateTelegraph}). Since all noise components are identical and uncorrelated, the autocorrelation of the total noise $\Delta(t)$ coincides with the one of a single telegraph noise $\llangle\Delta(0)\Delta(\tau)\rrangle=\sigma^2e^{-\kappa|\tau|}$, but higher order moments strongly depend on $M$. Due to the permutation symmetry of the dephasing environment, there are $M+1$ distinguishable realizations $\Delta_m$ of the total noise, labeled by $m=0,\dots,M$, and given by 
\begin{eqnarray}
\Delta_m=\frac{(2m-M)}{\sqrt{M}}\sigma.
\end{eqnarray} 
For instance, the realization $\Delta_0=-\sigma/\sqrt{M}$ corresponds to the configuration with all TLFs down, which vary all the way to $\Delta_M=\sigma/\sqrt{M}$ where all TLFs are up. A given realization $\Delta_m$ appears in the environment with a multiplicity ${{M}\choose{m}}=M!/[(M-m)!m!]$, and thus the probability $P(\Delta_m,t)$ to find the global realization $\Delta_m$ at time $t$ can be related to the probabilities of a single telegraph noise \changed{$P(\Delta_\pm,t)$} by 
\begin{eqnarray}
P(\Delta_m,t)={{M}\choose{m}}[P(\Delta_-,t)]^{M-m}[P(\Delta_+,t)]^{m},\qquad {\rm with} \ \ m=0,\dots, M.\label{probRelationM}
\end{eqnarray}
Using Eqs.~(\ref{MarkovianRateTelegraph}) and (\ref{probRelationM}), we can derive the rate equation for $P(\Delta_m,t)$, which takes the general form in Eq.~(\ref{RateEqDiscrete}), with a transition matrix $W_{nm}$ whose nonzero elements read \cite{hu2015},
\begin{eqnarray}
\fl\qquad\qquad W_{mm}=-\frac{M}{2} \kappa,\qquad W_{m,m+1}=\frac{\kappa}{2}(m+1),\qquad W_{m,m-1}=\frac{\kappa}{2}(M+1-m),\label{Mmulti}
\end{eqnarray}
for $m=0,\dots, M$, and the boundary conditions $P(\Delta_{-1},t)=P(\Delta_{M+1},t)=0$.

The steady state solution of the rate equation (\ref{RateEqDiscrete}) with the $W_{mn}$ coefficients (\ref{Mmulti}) is a binomial distribution ${P}_{\rm ss}(\Delta_m)=\frac{1}{2^M}{{M}\choose{m}}$ as can also be seen by setting ${P}_{\rm ss}(\Delta_\pm)=1/2$ in Eq.~(\ref{probRelationM}). Importantly, in the limit of an infinitely large ensemble of TLFs, $M\rightarrow \infty$, the binomial probability distribution ${P}_{\rm ss}(\Delta_m)$ approaches a continuous Gaussian distribution ${P}_{\rm G}(\Delta)=(2\pi\sigma^2)^{-1/2}e^{-\Delta^2/(2\sigma^2)}$ as ${P}_{\rm ss}(\Delta_m)={P}_{\rm G}(\Delta)d\Delta[1+{\cal O}(M^{-1/2})]$ and we recover the colored Gaussian noise limit of Secs.~\ref{IntrocoloredGaussian} and \ref{LineShapeColoredGaussian}. In fact, in the limit $M\rightarrow\infty$, the rate equation (\ref{RateEqDiscrete}) with (\ref{Mmulti}) becomes a continuous Fokker-Planck differential equation for the Ornstein-Uhlenbeck process \cite{hu2015}, which is given by Eqs.~(\ref{generalLiovillian})-(\ref{ChapmanKolmogorov}) \changed{with} $D_1(\Delta)=-\kappa\Delta$, $D_2=2\kappa\sigma^2$, and $W(\Delta,\Delta')=0$. As a result of this connection, we conclude that by increasing the number $M$ of independent telegraph noises, we can reduce the non-Gaussian character of the noise model until reaching the limit of standard colored Gaussian noise. 

We exemplify this tuning of the non-Gaussianity by computing the average transmittance $\llangle t_\omega^\mu\rrangle$ for a qubit in dephasing environments with different values of $M$. To do so, we numerically solve the linear system (\ref{LinearSystemEq})-(\ref{MmatrixLS}) by using the $W_{mn}$ coefficients in Eq.~(\ref{Mmulti}), and the steady state binomial distribution ${P}_{\rm ss}(\Delta_m)=\frac{1}{2^M}{{M}\choose{m}}$. It is computationally simple to reach the Gaussian limit $M\gg 1$ since the size of the matrix $J_{mn}$ grows linearly with $M$ as $(M+1)\times (M+1)$. The results are shown in Figure~\ref{fig:ResultsnonGaussianNoise}(b) for $M=[2,3,4,5,10]$, $\kappa=0.1\sigma$, and typical waveguide QED parameters. The non-Gaussianity of the dephasing is manifested by the multiple dips in $\llangle t_\omega^\mu\rrangle$ which reduce with increasing $M$. Also notice that already for $M=10$ (red/dashed) the Gaussian limit is well-established with a Gaussian-like transmittance as expected in the quasi-static limit $\kappa\ll\sigma<\infty$. In addition, we compute the Ramsey envelopes $C_\phi(t)$ for the parameters above by applying Eq.~(\ref{realRelation}) on the numerical data for $\llangle t_\omega^\mu\rrangle$. The results are shown in Figure~\ref{fig:ResultsnonGaussianNoise}(c), where the non-Gaussinity of the dephasing noise is manifested by the multiple oscillations in $C_\phi(t)$ and whose amplitude reduce with $M$. In the Gaussian limit (red/dashed) there is only the Gaussian decay as expected in the quasi-static case $\kappa=0.1\sigma$. Notice that we do not display the results in the white noise limit, where the behavior is independent of $M$, the lineshapes are standard Lorentzians, and $C_\phi(t)$ are exponential decays with pure dephasing rate $\gamma_\phi=\sigma^2/\kappa$.

\subsection{Simulation of non-Gaussian 1/f noise}\label{sec:1fnoise}

The aim of this subsection is to construct a model for 1/f noise with tunable non-Gaussianity and show how to compute the non-Gaussian results for $\llangle t_\omega^\mu\rrangle$ in Figure~\ref{fig:Gaussian1f}(b)-(c). To similate non-Gaussian 1/f noise, we assume that each noise component $\Delta_j(t)$ for $j=1,\dots, N$ in Eq.~(\ref{uncorrelatedSum}) is represented by an independent ensemble of $M$ identical TLFs as introduced in~\ref{sec:tunableGaussian}. We therefore need to construct a more general jump model for the total noise, $\Delta(t)=\sum_{j=1}^N \Delta_j(t)/\sqrt{N}$, with permutation symmetry only within each ensemble $\Delta_j(t)$. As a result, there will be $(M+1)^N$ distinguishable global realizations of the total noise $\Delta(t)$, which are given by
\begin{eqnarray}
\Delta_{\vec{m}}=\sum_{j=1}^N \frac{(2m_j-M)}{\sqrt{M}}\sigma_j.\label{VectorRealizations}
\end{eqnarray}
Here, we use the vectorial index $\vec{m}=(m_1,\dots,m_N)$, with components $m_j=0,\dots,M$, to label the above $(M+1)^N$ different realizations $\Delta_{\vec{m}}$. Then, by straightforwardly generalizing the procedure in~\ref{sec:tunableGaussian}, one can show that probability $P(\Delta_{\vec{m}},t)$ satisfies a rate equation of the form (\ref{RateEqDiscrete}), with a transition matrix $W_{\vec{m}\vec{n}}$ of size $(M+1)^N\times (M+1)^N$ and nonzero matrix elements given by,
\begin{eqnarray}
\fl\quad W_{\vec{m}\vec{m}}=-\frac{M}{2}\sum_{j=1}^N \kappa_j,\qquad W_{\vec{m},\vec{m}+\vec{e}_j}=\frac{\kappa_j}{2}(m_j+1),\qquad W_{\vec{m},\vec{m}-\vec{e}_j}=\frac{\kappa_j}{2}(M+1-m_j),
\end{eqnarray}
where $\vec{e}_j=(0,\dots, 1_j,\dots, 0)$ is a unit vector in component $j=1,\dots,N$. Solving the corresponding rate equation with boundary conditions ${P}(\Delta_{\vec{m}},t)=0$ for $m_j=-1,M+1$, and $j=1,\dots, N$, we find that the steady state probability ${P}_{\rm ss}(\Delta_{\vec{m}})$ corresponds to a product of binomial distributions for each $\Delta_j(t)$, which reads
\begin{eqnarray}
{P}_{\rm ss}(\Delta_{\vec{m}})=\frac{1}{2^{NM}}\prod_{j=1}^N {M\choose m_j}.\label{VectorProb}
\end{eqnarray}

Finally, we should evaluate $W_{\vec{m}\vec{n}}$ for the parameters $\kappa_j$ and $\sigma_j$ that simulate the desired 1/f noise model as stated in Sec.~\ref{LineShape1f}, replace this and Eq.~(\ref{VectorProb}) in the linear system (\ref{LinearSystemEq})-(\ref{MmatrixLS}), and numerically solve for the steady state marginal averages. With that result we can evaluate $\llangle t_\omega^\mu\rrangle$ via Eq.~(\ref{TransmittanceDiscrete}), and $C_\phi(t)$ via Eq.~(\ref{realRelation}). The size of the linear system scales exponentially with $N$ as $(M+1)^N$, but as shown in Figure~\ref{fig:Gaussian1f}(a), already a moderate $N=8$ is enough to properly simulate the 1/f noise spectrum.

\section{Correlated dephasing noise in a qubit with Fano resonance}\label{FanoRelations}

Some waveguide QED experiments are affected by input-output impedance mismatches or internal reflections that impose a Fano resonance profile on the scattering experiment \cite{thyrrestrup17,javadi15}. We briefly discuss how to modify our protocol \changed{and the scattering equations} for reconstructing the power measurements and the noise correlations in such complex environments.

Following Refs.~\cite{auffeves07,thyrrestrup17,javadi15}, \com{we see that a Fano resonance can be modeled by a highly dissipative cavity mode that mediates the coupling between the propagating photons and the qubit. In this case, the cavity mode can be adiabatically eliminated \cite{auffeves07} and the effective dynamics of the qubit is governed by quantum Langevin equations} with the same form as Eqs.~(\ref{EqMotSm})-(\ref{EqMotSz}), but with a modified total decay $\Gamma\rightarrow \gamma_{\rm loss}+{\rm Re}\lbrace z_\omega \rbrace \gamma$, a modified qubit central frequency $\omega_0\rightarrow \omega_0+{\rm Im}\lbrace z_\omega \rbrace \gamma/2$, and a modified input operator $a_{\rm in}^\mu(t)\rightarrow z_\omega a_{\rm in}^\mu(t)$. The correction $z_\omega$ is the Fano resonance function\changed{, which depends on the frequency of the incident photon $\omega$ and is} given by
\begin{eqnarray}
z_\omega=\frac{1}{1-2i(\omega-\omega_c)/\kappa},
\end{eqnarray}
with $\omega_c$ the resonance frequency and $\kappa$ the decay of the localized mode producing the Fano resonance. In addition, the input-output relations (\ref{InputOutput}) are modified as \cite{auffeves07} 
\begin{eqnarray}
a_{\rm out}^\mu(t) = \sum_\lambda \Lambda_{\mu\lambda}(\omega)a_{\rm in}^\lambda(t)+iz_\omega\sqrt{\gamma_\mu}\sigma^-(t),
\end{eqnarray}
with coefficients $\Lambda_{\mu\lambda}(\omega) = \delta_{\mu\lambda} - 2 z_\omega \sqrt{\gamma_\mu \gamma_\lambda}/\gamma$, and the indices $\mu,\lambda=\pm$ corresponding to photons propagating to the right $(+)$ and left $(-)$ of the waveguide.

\subsection{Single-photon scattering matrix of a noisy qubit with Fano resonance}\label{FanoMatrix}

\changed{From the modified Langevin equations and input-output relation stated above, we can calculate the average single-photon scattering matrix $\llangle S_{\nu\omega}^{\lambda\mu}\rrangle_{\rm Fano}$ for a qubit with Fano resonance, using the same procedure and definitions shown in Sec.~(\ref{averageSMatrix}). We obtain,}
\begin{eqnarray}
\llangle S^{\lambda\mu}_{\nu\omega}\rrangle_{\rm Fano} = \left\lbrace \Lambda_{\mu\lambda}(\omega)+z_\omega\sqrt{\gamma_\lambda\gamma_\mu}\llangle G_{\omega}\rrangle_{\rm Fano}\right\rbrace\delta(\nu-\omega),
\end{eqnarray}
with
\begin{eqnarray}
\llangle G_\omega\rrangle_{\rm Fano} = {\cal L}[C_\phi(t)]\left([z_\omega \gamma+\gamma_{\rm loss}]/2-i[\omega-\omega_0]\right).\label{GFano}
\end{eqnarray}
The average single-photon transmittance and reflectance in the presence of correlated noise then read,
\begin{eqnarray}
\fl\llangle t_\omega^\mu\rrangle_{\rm Fano} =  1-\frac{z_\omega\gamma_\mu}{\gamma/2}+z_\omega \gamma_\mu \llangle G_\omega\rrangle_{\rm Fano},\qquad \llangle r_\omega^\mu\rrangle_{\rm Fano} = -\frac{z_\omega\sqrt{\gamma_+\gamma_-}}{\gamma/2}\left(1-\frac{\gamma}{2} \llangle G_\omega\rrangle_{\rm Fano}\right).\label{FanoSinglePhoton}
\end{eqnarray}
Notice that in the case of an exact Fano resonance ($\omega_c=\omega$), the qubit effectively behaves as it would be directly coupled to two independent waveguides on each side as treated in Refs.~\cite{shen09new,roy17}. This situation is known as a ``direct-coupled'' qubit in contrast to the ``side-coupled'' qubit we consider throughout the main text. It is discussed in Refs.~\cite{shen09new,roy17} that the results of both cases are related, up to a phase, by interchanging the roles of transmission and reflection. Here, by setting $z_\omega=1$ in Eqs.~(\ref{FanoSinglePhoton}), and considering a non-chiral case $\gamma_\mu=\gamma/2$, we find that these relations are still valid in the presence of correlated noise, namely $\llangle t_\omega^\mu\rrangle_{\rm Fano}=-\llangle r_\omega^\mu\rrangle$, and $\llangle r_\omega^\mu\rrangle_{\rm Fano}=-\llangle t_\omega^\mu\rrangle$.

\subsection{\changed{Power and homodyne measurements of a noisy qubit with Fano resonance}}\label{FanoReconstruction}

\changed{Using the replacements $\Gamma\rightarrow \gamma_{\rm loss}+{\rm Re}\lbrace z_\omega \rbrace \gamma$, $\omega_0\rightarrow \omega_0+{\rm Im}\lbrace z_\omega \rbrace \gamma/2$, and $\Omega\rightarrow z_\omega \Omega$ in the optical Bloch equations (\ref{BlochSigmamin})-(\ref{BlochSigmaZ}), we can generalize Eqs.~(\ref{TMeasurementHomodyne})-(\ref{KK}) and (\ref{reflectionhomodyne})-(\ref{reflectionKK}) for the homodyne or power measurements, and obtain 
\begin{eqnarray}
\frac{\llangle\braket{a^\lambda_{\rm out}}\rrangle_{\rm ss}}{\alpha_\omega^\mu} = \Lambda_{\mu\lambda}(\omega)+z_\omega^2 \sqrt{\gamma_\mu\gamma_\lambda}\ \llangle Q_\omega\rrangle,\label{homodyneFano}\\
\frac{\llangle \braket{a^\lambda_{\rm out}{}^\dag a^\lambda_{\rm out}}\rrangle_{\rm ss}}{|\alpha_\omega^\mu|^2} = |\Lambda_{\mu\lambda}(\omega)|^2 +2\sqrt{\gamma_\mu\gamma_\lambda}\ {\rm Re}\left\lbrace {\cal K}_{\mu\lambda}(\omega) \llangle Q_\omega\rrangle\right\rbrace,\label{powerFano}\\
{\rm Im}\left\lbrace {\cal K}_{\mu\lambda}(\omega) \llangle Q_\omega\rrangle \right\rbrace = -\frac{1}{\pi}{\cal P}\int_{-\infty}^\infty d\omega' \frac{{\rm Re}\left\lbrace {\cal K}_{\mu\lambda}(\omega') \llangle Q_{\omega'}\rrangle \right\rbrace}{\omega'-\omega}.
\end{eqnarray}
Here, $\llangle Q_\omega\rrangle=\llangle \sigma^-\rrangle_{\rm ss}/\Omega$, and the coefficients ${\cal K}_{\mu\lambda}(\omega)$ read
\begin{eqnarray} 
{\cal K}_{\mu\lambda}(\omega) = z_\omega^2 \Lambda_{\mu\lambda}^\ast(\omega) + \frac{|z_\omega|^4 \sqrt{\gamma_\mu \gamma_\lambda}}{\left(|z_\omega|^2\gamma + \gamma_{\rm loss}\right)}.
\end{eqnarray}
The new equations (\ref{homodyneFano})-(\ref{powerFano}) are valid for measuring at both the transmission ($\lambda=\mu$) and the reflection ($\lambda=-\mu$) output, and provide a robust method to infer $\llangle Q_\omega\rrangle$, which is related to the average scattering overlap $\llangle G_\omega\rrangle_{\rm Fano}$ in Eq.~(\ref{GFano}) as
\begin{eqnarray}
\llangle Q_\omega\rrangle = \llangle G_\omega\rrangle_{\rm Fano}+{\cal O}\left[|\Omega|/\Gamma\right]^2,\label{relationFano}
\end{eqnarray}
in the limit $|\Omega|\ll \Gamma$. Using Eqs.~(\ref{powerFano})-(\ref{relationFano}) we can experimentally determine $\llangle G_\omega\rrangle_{\rm Fano}$ and from there obtain the single-photon transmission and reflection coefficients (\ref{FanoSinglePhoton}), in the case of a Fano resonance. Finally, from the knowledge of $\llangle G_\omega\rrangle_{\rm Fano}$ we can also invert Eq.~(\ref{GFano}), in analogy to Eq.~(\ref{realRelation}), and recover the Ramsey profile from the above spectroscopic measurements as
\begin{eqnarray}
C_\phi(t)=\frac{1}{2\pi}e^{\gamma_{\rm loss}t/2}{\cal F}^{-1}\left[e^{z_\omega\gamma t/2}\llangle G_\omega\rrangle_{\rm Fano}\right](t),\quad {\rm for}\quad t>0,
\end{eqnarray}
where we can use ${\cal F}^{-1}$ instead of ${\cal L}^{-1}$ due to the non-zero emission rates into guided $\gamma$ or unguided $\gamma_{\rm loss}$ modes.}

\section{Adding a white noise background to the dephasing model}\label{sec:WNbackground}

In this appendix, we use stochastic Ito calculus \cite{jacobs2010,gardiner1985} to include dephasing due to a white noise background $\Delta_{\rm WB}(t)$, in addition to the correlated noise $\Delta(t)$ in the scattering differential equation (\ref{ScatteringEq}).

The stochastic differential equation for scattering that includes both noise sources reads,
\begin{eqnarray}
\frac{d}{dt}G_{\omega}(t)=&-\left(\frac{\Gamma}{2}-i[\omega-\omega_0]+i[\Delta(t)+\Delta_{\rm WB}(t)]\right)G_{\omega}(t)+1,\label{stochasticScatteringWB}
\end{eqnarray}
where the white noise background is specified by the autocorrelation function $\llangle \Delta_{\rm WB}(0)\Delta_{\rm WB}(\tau)\rrangle=2\gamma_{\rm WB}\delta(\tau)$, with $\gamma_{\rm WB}$ its pure dephasing rate. The multiplicative stochastic differential equation (\ref{stochasticScatteringWB}) must be physically interpreted in the Stratonovich form \cite{jacobs2010,gardiner1985},
\begin{eqnarray}
\fl({\rm S})\quad d G_{\omega}(t)=&-\left(\frac{\Gamma}{2}-i[\omega-\omega_0]+i\Delta(t)\right)G_{\omega}(t)dt+dt+i\sqrt{2 \gamma_{\rm WB}}G_{\omega}(t)dW(t),\label{stratonovich}
\end{eqnarray} 
with $dW(t)=\Delta_{\rm WB}(t)dt/\sqrt{2\gamma_{\rm WB}}$ the Wiener increment. To solve the average over the white noise background more easily, we use the Ito rules to convert Eq.~(\ref{stratonovich}) to the Ito form, obtaining
\begin{eqnarray}
\fl({\rm I})\quad d G_{\omega}(t)=&-\left(\frac{\Gamma}{2}+\gamma_{\rm WB}-i[\omega-\omega_0]+i\Delta(t)\right)G_{\omega}(t)dt+dt+i\sqrt{2 \gamma_{\rm WB}}G_{\omega}(t)dW(t),
\end{eqnarray}  
where now $dW(t)$ is uncorrelated with $G_{\omega}(t)$ at equal times. We take the average over the white noise background $\llangle\dots\rrangle_{\rm WB}$, which does not affect $\Delta(t)$ as we assume it is uncorrelated with $\Delta_{\rm WB}(t)$, i.e.~$\llangle \Delta(t) \Delta_{\rm WB}(t)\rrangle_{\rm WB}=0$ and $\llangle \Delta(t) G_{\omega}(t)\rrangle_{\rm WB}=\Delta(t) \llangle G_{\omega}(t)\rrangle_{\rm WB}$. Additionally using the Ito property $\llangle G_{\omega}(t) dW(t)\rrangle_{\rm WB}=\llangle G_{\omega}(t)\rrangle_{\rm WB} \llangle dW(t)\rrangle_{\rm WB}=0$, we obtain a stochastic differential equation that depends on the correlated noise $\Delta(t)$ only,
\begin{eqnarray}
\frac{d}{dt}\llangle G_{\omega}\rrangle_{\rm WB}=-\left(\frac{\Gamma}{2}+\gamma_{\rm WB}-i[\omega-\omega_0]+i\Delta(t)\right)\llangle G_{\omega} \rrangle_{\rm WB}(t)+1.
\end{eqnarray}
Therefore, we can solve this stochastic differential equation instead of (\ref{ScatteringEq}) if we would like to include an extra uncorrelated white noise background with pure dephasing rate $\gamma_{\rm WB}$. In practice it just amounts to perform the replacement $\Gamma/2\rightarrow\Gamma/2+\gamma_{\rm WB}$ in Eq.~(\ref{ScatteringEq}), before starting to solve it.

\section{Measurement of single-photon reflectance and conservation of average photon flux}\label{ReflectionExpressions}

In this appendix we complement the analysis from section~\ref{sec:coherentextraction}, providing formulas for the average reflectance $\llangle r_\omega^{\mu}\rrangle$, and a word of caution on the interpretation of the squares of the averages $|\llangle r_\omega^{\mu}\rrangle|^2$ and $|\llangle t_\omega^{\mu}\rrangle|^2$, in the presence of dephasing.

The average single-photon transmittance $\llangle t_\omega^{\mu}\rrangle$ can be measured via Eqs.~(\ref{TMeasurementHomodyne})-(\ref{KK}) in Sec.~\ref{sec:coherentextraction} when performing homodyne or power measurements at the output of the same channel $\mu=\pm$ as the weak input drive $\alpha_\omega^\mu$. If we instead perform the measurements at the opposite channel $\lambda=-\mu$, we access to the average reflectance $\llangle r_\omega^{\mu}\rrangle$ via the relations,
\begin{eqnarray}
\frac{\llangle \braket{\tilde{a}^{(-\mu)}_{\rm out}(t)} \rrangle_{\rm ss}}{\alpha_\omega^\mu}=\llangle r_\omega^{\mu}\rrangle + {\cal O}\left[|\Omega|/\Gamma \right]^2,\label{reflectionhomodyne}\\
\frac{\llangle \braket{a^{(-\mu)}_{\rm out}{}^\dag(t)a^{(-\mu)}_{\rm out}(t)}\rrangle_{\rm ss}}{|\alpha_\omega^\mu|^2} =  -2\sqrt{\beta_\lambda\beta_\mu}{\rm Re}\lbrace \llangle r_\omega^{\mu}\rrangle\rbrace+{\cal O}\left[|\Omega|/\Gamma\right]^2,\\
{\rm Im}\lbrace \llangle r_\omega^{\mu} \rrangle \rbrace = -\frac{1}{\pi}{\cal P}\int_{-\infty}^\infty d\omega' \frac{{\rm Re}\lbrace \llangle r_{\omega'}^{\mu} \rrangle \rbrace}{\omega'-\omega}.\label{reflectionKK}
\end{eqnarray}

When a quantum emitter is affected by dephasing, the squares of the average transmittance and reflectances do not add to one. \com{This is because the dephasing environment exerts work, adding and subtracting energy on the qubit in order to change its transition frequency. For stationary noise the average work is zero, but still the system of qubit and photons is open due to the external stochastic field $\Delta(t)$.} In the simple case of white noise dephasing, we can evaluate Eqs.~(\ref{AverageReflectance}) and (\ref{LorentzianTransmittanceWN}) and obtain
\begin{eqnarray}
|\llangle t_\omega^{\mu} \rrangle|^2+|\llangle r_\omega^{\mu} \rrangle|^2+|\llangle r_\omega^{\mu,{\rm loss}} \rrangle|^2=1-\frac{\gamma_\phi\gamma_\mu}{(\Gamma/2+\gamma_\phi)^2+(\omega-\omega_0)^2},\label{coefficientsSquare}
\end{eqnarray}
with $\llangle r_\omega^{\mu,{\rm loss}} \rrangle=\sqrt{\gamma_{{\rm loss}}/\gamma_\mu}\left(\llangle t_\omega^\mu\rrangle-1\right)$ the fluorescence reflectance into unguided modes, and $\gamma_\phi$ the pure dephasing rate.

This means that the squares of the average transmittance or reflectance do not describe photon fluxes when $\gamma_\phi\neq0$. The noisy qubit indeed conserves the total photon flux on average, in the case of stationary dephasing, but this is manifested in the sum of the average output power in all channels, i.e.~transmission, reflection, and fluorescence loss, as
\begin{eqnarray}
\fl\qquad\ \ \frac{\llangle \braket{a^{\mu}_{\rm out}{}^\dag(t) a^{\mu}_{\rm out}(t)}\rrangle_{\rm ss}}{|\alpha_\omega^\mu|^2}+\frac{\llangle \braket{a^{(-\mu)}_{\rm out}{}^\dag(t) a^{(-\mu)}_{\rm out}(t)}\rrangle_{\rm ss}}{|\alpha_\omega^\mu|^2}+\frac{\llangle \braket{a^{\rm loss}_{\rm out}{}^\dag(t) a^{\rm loss}_{\rm out}(t)}\rrangle_{\rm ss}}{|\alpha_\omega^\mu|^2}=1.\label{powerConservationTotal}
\end{eqnarray}

\end{document}